\RequirePackage[T1]{fontenc}
\documentclass[11pt, urlcolor=blue, linkcolor=blue]{article} 
\usepackage{cite}
\usepackage{float}
\usepackage[utf8]{inputenc}

\usepackage{multicol}
\usepackage{multirow}
\usepackage{tablefootnote}
\usepackage{amsmath, amsthm, amssymb,slashed,mathtools,tabu}
\usepackage{ifpdf}
\ifpdf
  \usepackage[pdftex]{graphicx}
  \usepackage{epstopdf}
\else
  \usepackage[dvips]{graphicx}
\fi
\parskip=\baselineskip

\usepackage[usenames, dvipsnames]{color}
\usepackage[svgnames]{xcolor}
\usepackage[colorlinks,citecolor=RoyalBlue, urlcolor=RoyalBlue, linkcolor=RoyalBlue ]{hyperref} 

\allowdisplaybreaks[1]

\numberwithin{equation}{section}

\usepackage{booktabs}

\theoremstyle{definition}

\newcommand{\cblue}[1]{\textcolor{blue}{#1}}
\newcommand{\cred}[1]{\textcolor{black}{#1}}

\definecolor{mygray}{gray}{0.6}

\usepackage{upgreek}
\usepackage{bbm}

\newenvironment{myfont}[2][]{\csname#2\endcsname[#1]}{}

\usepackage{slashed}
\usepackage[makeroom]{cancel}
\usepackage[normalem]{ulem}
\usepackage{soul}
\newcommand{\stkout}[1]{\ifmmode\text{\sout{\ensuremath{#1}}}\else\sout{#1}\fi}

\usepackage{sseq}
\usepackage[all,cmtip]{xy}
\usepackage{tikz-cd}
\usepackage{tikz}
\usetikzlibrary{matrix}

\newcommand{\bea}{\begin{eqnarray}}
\newcommand{\eea}{\end{eqnarray}}
\def\be{\begin{equation}}
\def\ee{\end{equation}}

\newcommand{\e}{\hspace{1pt}\mathrm{e}}

\newcommand{\ii}{\hspace{1pt}\mathrm{i}\hspace{1pt}}

\definecolor{red}{rgb}{1,0,0}
\definecolor{blue}{rgb}{0,0,1}
\definecolor{dblue}{rgb}{0,0,0.4}
\definecolor{green}{rgb}{0,1,0}
\definecolor{black}{rgb}{0,0,0}
\definecolor{white}{rgb}{1,1,1}

\definecolor{brn}{rgb}{.8,.4,.0}
\definecolor{redo}{rgb}{1,.5,.0}
\definecolor{ddgrn}{rgb}{0,0.4,0}
\definecolor{dgrn}{rgb}{0,0.55,0}
\definecolor{dbl}{rgb}{0,0,0.5}

\usepackage[bbgreekl]{mathbbol}
\usepackage{amscd}

\newcommand{\Z}{\mathbb{Z}}
\newcommand{\C}{\mathbb{C}}

\newcommand{\dd}{\hspace{1pt}\mathrm{d}}

\newcommand{\Refe}[1]{Ref.~\cite{#1}}
\newcommand{\Eq}[1]{(\ref{#1})} 
\newcommand{\eq}[1]{(\ref{#1})} 
 
\newcommand{\Eqn}[1]{Eqn.~(\ref{#1})} 

\newcommand{\Tr}{{\rm Tr}}

\newcommand{\bpm}{\begin{pmatrix}}
\newcommand{\epm}{\end{pmatrix}}
\newcommand{\bmm}{\begin{matrix}}
\newcommand{\emm}{\end{matrix}}

\newcommand{\cA}{ {\cal A} } 
\newcommand{\cB}{ {\cal B} }

\newcommand{\cL}{ {\cal L} }

\def\CA{{\cal A}}

\def\CX{{\cal X}}

\def\Z{{\mathbb{Z}}}

\def\C{{\mathbb{C}}}

\def\Tr{{\mathrm{Tr}}}

\def \H{\operatorname{H}}

\def \Z{\mathbb{Z}}
\def \Pin{\mathrm{Pin}}

\newcommand{\Sec}[1]{Sec.~\ref{#1}}

\usetikzlibrary{decorations.markings}
\usetikzlibrary{positioning}
\usetikzlibrary{shadings}

\newcommand{\SO}{{\rm SO}}
\newcommand{\Spin}{{\rm Spin}}
\newcommand{\U}{{\rm U}}
\newcommand{\SU}{{\rm SU}}

\renewcommand{\O}{{\rm O}}

\newcommand{\rS}{{\rm S}}
\newcommand{\rT}{{\rm T}}

\newcommand{\rF}{{\rm F}}
\newcommand{\rN}{{\rm N}}

\def\Sq{\mathrm{Sq}}

\def\B{\mathrm{B}}

\def\TP{\mathrm{TP}}

\usepackage{datetime}
\usepackage{enumitem} 
\usepackage{moreenum}

\newcommand{\sharpfootnote}[1]{%
\let\oldthefootnote=\thefootnote%
\stepcounter{mpfootnote}%
\addtocounter{footnote}{-1}%
\renewcommand{\thefootnote}{{W}} 
\footnote{#1}%
\let\thefootnote=\oldthefootnote%
}

\newcommand{\naturalfootnote}[1]{%
\let\oldthefootnote=\thefootnote%
\stepcounter{mpfootnote}%
\addtocounter{footnote}{-1}%
\renewcommand{\thefootnote}{{W$^-\natural$}}
\footnote{#1}%
\let\thefootnote=\oldthefootnote%
}

\newcommand{\flatfootnote}[1]{%
\let\oldthefootnote=\thefootnote%
\stepcounter{mpfootnote}%
\addtocounter{footnote}{-1}%
\renewcommand{\thefootnote}{{W$^-\flat$}}
\footnote{#1}%
\let\thefootnote=\oldthefootnote%
}

\DeclareRobustCommand\sWang
{\cblue{\includegraphics[height=3.2ex]{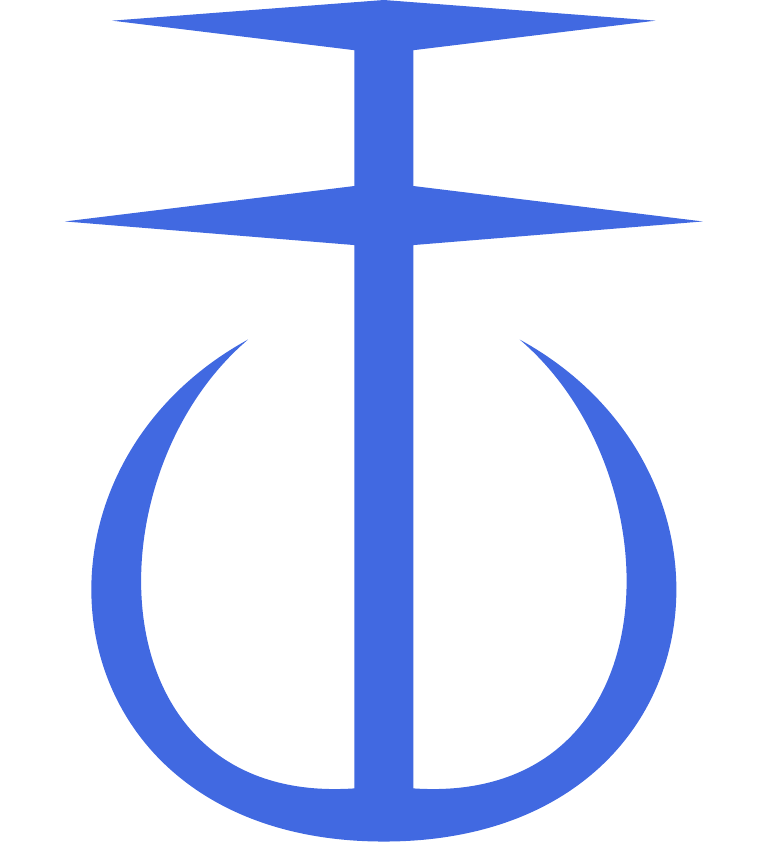}}}

\newcommand{\Wangfootnote}[1]{%
\let\oldthefootnote=\thefootnote%
\stepcounter{mpfootnote}%
\addtocounter{footnote}{-1}%
\renewcommand{\thefootnote}{\sWang}
\footnote{#1}%
\let\thefootnote=\oldthefootnote%
}

\usepackage{comment}
\def\bZ{{\mathbf{Z}}}

\usepackage{upgreek}
\newcommand{\Fig}[1]{Fig.~\ref{#1}}

\newcommand{\SM}{{\rm SM}} 
\newcommand{\PD}{{\rm PD}} 
\newcommand{\UU}{{\rm UU}} 
\newcommand{\GUT}{{\rm GUT}} 
 
\newcommand{\TQFT}{{\rm TQFT}} 
\newcommand{\ABK}{\text{ABK}}

\def\ra{\mathrm{a}}

\def\rc{\mathrm{c}}

\usepackage{mathrsfs}
\usepackage{esint} 

\begin{document}
\begin{titlepage}
\begin{flushright}
\end{flushright}
\begin{center}


{\bf\LARGE{ 
Ultra Unification
}}

\vskip0.5cm 
\Large{Juven Wang
\Wangfootnote{
{\tt jw@cmsa.fas.harvard.edu} 
\quad\quad\quad\quad\quad\quad\quad\quad\;\; 
\includegraphics[width=1.0in]{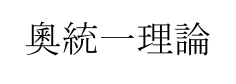}
\quad\quad \hfill 
\href{http://idear.info/}{Talk slides download:}
\href{http://idear.info/}{\bf http://idear.info/}
\\[2mm]
\quad\quad\quad\quad \quad\quad  Physical Review D version subtitle: Unified model beyond grand unification. 
\quad\quad \href{https://www.youtube.com/results?search_query=ultra+unification+Quantum+Criticality+juven+wang+deformation+standard+model}{Seminar presentation 
videos available online}
} 
\\[2.75mm]  
} 
\vskip.5cm
{ {\small{\textit{
{Center of Mathematical Sciences and Applications, Harvard University,  Cambridge, MA 02138, USA}}}}
}

\end{center}

\vskip 0.5cm
\baselineskip 16pt
\begin{abstract}

Strong, electromagnetic, and weak forces were unified in the Standard Model (SM) with spontaneous gauge \emph{symmetry breaking}. 
These forces were further conjectured to be unified in a simple Lie group gauge interaction in the Grand Unification (GUT).
In this work, we propose a theory 
beyond the SM and 
GUT 
by adding new gapped Topological Phase Sectors consistent with the nonperturbative global anomaly cancellation 
and cobordism constraints (especially from the baryon minus lepton number ${\bf B}-{\bf L}$, the electroweak hypercharge $Y$,
and the mixed gauge-gravitational anomaly). 
Gapped Topological Phase Sectors are constructed via \emph{symmetry extension},
whose low energy contains unitary Lorentz invariant topological quantum field theories (TQFTs):
either 3+1d non-invertible TQFT (long-range entangled gapped phase), 
or 4+1d invertible or non-invertible TQFT (short-range or long-range entangled gapped phase). Alternatively,
there could also be right-handed ``sterile'' neutrinos, gapless unparticle physics, more general 
interacting conformal field theories, or gravity with topological cobordism constraints, 
or their combinations to altogether cancel the mixed gauge-gravitational anomaly.
We propose that a new high-energy physics frontier beyond the 
conventional 0d particle physics 
relies on the new Topological Force
and Topological Matter including gapped extended objects 
(gapped 1d line and 2d surface operators or defects, etc., whose open ends carry deconfined fractionalized particle or anyonic string excitations)
or gapless conformal matter. 
Physical characterizations of these gapped extended objects require the 
mathematical theories of cohomology, cobordism, or category.
Although weaker than the weak force, 
Topological Force is infinite-range or long-range which does not decay in the distance, and mediates between
the linked worldvolume trajectories via fractional or categorical statistical interactions.\\[10mm]


%
%


\flushright
December 2020

\end{abstract}

\end{titlepage}

  \pagenumbering{arabic}
    \setcounter{page}{2}
    

\tableofcontents

\newpage
~\newline ~\newline ~\newline ~\newline ~\newline ~\newline
~\newline ~\newline
~\newline

\hfill ``Die einzig wahre Unsterblichkeit liegt in den eigenen Kindern.'' \\[1mm] 
\hspace*{\fill}``The only true immortality resides in one's children.'' \\[2mm] 

\hspace*{\fill}``Meine Arbeit ist Ihnen gewidmet. Sie k\"onnen dies sicherlich in Ihrem Herzen f\"uhlen.'' \\[1mm] 
\hspace*{\fill}``My work is dedicated to you. You can surely feel this in your heart.'' \\[2mm] 

\hspace*{\fill} Intermezzo in A major. Andante teneramente, Six Pieces for Piano, Op.~118\\[1mm] 
\hspace*{\fill} Johannes Brahms in 1893

\newpage
\section{Introduction and Summary}
Unification is a central theme in theoretical physics. 
In 1864-1865, Maxwell  \cite{Maxwell1865zz}
unified the electricity and magnetism into the electrodynamics theory, where 
the derived electromagnetic wave manifests the light phenomena.
In 1961-1967,
Glashow-Salam-Weinberg (GSW) \cite{Glashow1961trPartialSymmetriesofWeakInteractions, Salam1964ryElectromagneticWeakInteractions, Salam1968, Weinberg1967tqSMAModelofLeptons} 
made landmark 
contributions to the electroweak theory of the unified electromagnetic and weak forces 
between elementary particles, including the prediction of the weak neutral current.
The GSW theory together with the strong force \cite{Gross1973id, Politzer1973fx} is now known as the Standard Model (SM), 
which is verified to be theoretically and experimentally essential to 
describe the subatomic high energy physics (HEP). 
In 1974, Georgi-Glashow
hypothesized that at a higher energy, 
the three gauge interactions of the SM 
would be merged into a single electronuclear force
under a simple Lie group gauge theory, known as the Grand Unification or Grand Unified Theory (GUT) \cite{Georgi1974syUnityofAllElementaryParticleForces, Fritzsch1974nnMinkowskiUnifiedInteractionsofLeptonsandHadrons}.

\begin{figure}[h!] 
  \centering
  \hspace{-1.85cm}
  \;\includegraphics[width=7.4in]{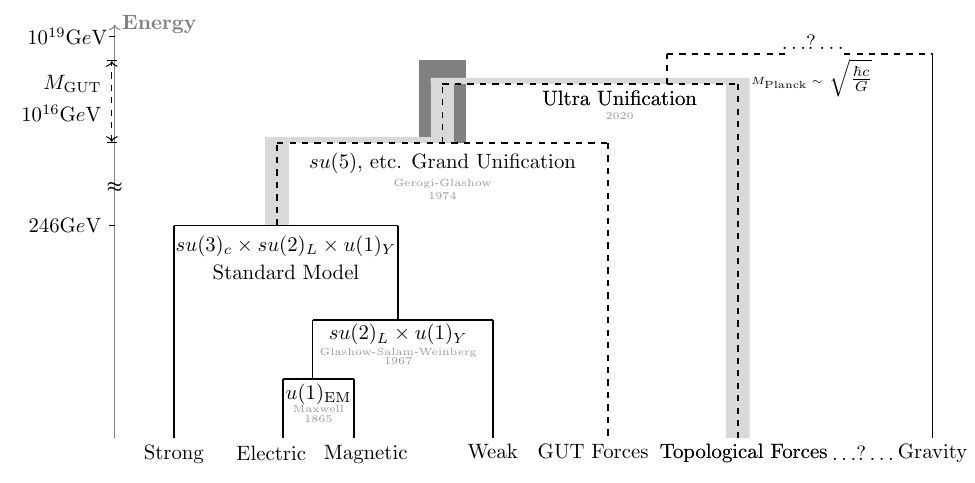}
  \caption{Unification of forces and interactions in fundamental physics.
  In fact, the proposed Ultra Unification can still play a role of unification even if Grand Unification (GUT) is not favored by Nature or verified by experiments.
  Namely, based on the anomaly and cobordism constraint, 
  we can still propose the Standard Model (SM) + gapped Topological Phase Sector or gapless conformal sector + Topological Force in a consistent way without GUT.
  The \emph{dark} gray area schematically shows the possible energy scales for various GUT scenarios, such as the $su(5), so(10), \dots, so(18)$ GUT around $10^{16}$G$e$V.
  The \emph{light} gray area schematically suggests that the possible energy gap $\Delta_{\rm TQFT}$ for Topological Phase Sector can range from
  as low energy as the SM, to as high energy to somewhere within the GUT scales (e.g., below the $so(10)$ GUT scale). 
  The dashed lines mean hypothetical unifications that have not yet been confirmed by experiments. 
  Forces are arranged from the strongest to the weakest (horizontally from the left to the right) in the electroweak Higgs vacuum.
  }
  \label{fig:BSM-UU-tree}
\end{figure}

In this work,
following our previous investigations based on nonperturbative global anomalies and cobordism constraints 
\cite{WW2019fxh1910.14668, JW2006.16996,JW2008.06499},\footnote{Related global anomalies and cobordism constraints on SM, GUT, and BSM
are explored also in \cite{Freed0607134, GarciaEtxebarriaMontero2018ajm1808.00009, WangWen2018cai1809.11171, DavighiGripaiosLohitsiri2019rcd1910.11277}.
Freed gave a generalized cohomology description of the $su(5)$ GUT anomalies \cite{Freed0607134}. 
Garcia-Etxebarria-Montero \cite{GarciaEtxebarriaMontero2018ajm1808.00009} 
and Davighi-Gripaios-Lohitsiri \cite{DavighiGripaiosLohitsiri2019rcd1910.11277} used Atiyah-Hirzebruch spectral sequence (AHSS) to compute the classification of global anomalies.
The author with Wen \cite{WangWen2018cai1809.11171} and with Wan \cite{WW2019fxh1910.14668} used Adams spectral sequence (ASS) \cite{Adams1958},
Thom-Madsen-Tillmann spectra \cite{thom1954quelques,MadsenTillmann4}, and Freed-Hopkins theorem \cite{Freed2016}
to obtain the classification of all invertible quantum anomalies. In addition, there in \Refe{WW2019fxh1910.14668},
we can also fully characterize the $d$d 't Hooft anomaly \cite{tHooft1979ratanomaly}
of global symmetry $G$
as $(d+1)$d cobordism invariants, precisely as topological terms of cohomology classes and fermionic topological invariants.
The cobordism invariants can be read from the Adams chart in \Refe{WW2019fxh1910.14668}.
Thus we focus on employing \Refe{WW2019fxh1910.14668} result.
By classifying all the invertible quantum anomalies,
we must include 
\begin{itemize}[leftmargin=-0.mm]
\item all \emph{local anomalies} (\emph{perturbative anomalies}):\\ 
captured by perturbative Feynman diagram loop calculations, 
classified by the integer $\mathbb{Z}$ classes (the free classes).\\ 
e.g., Adler-Bell-Jackiw (ABJ) anomalies \cite{Adler1969gkABJ,Bell1969tsABJ}, perturbative
{local} gravitational anomalies\cite{AlvarezGaume1983igWitten1984}.
{Typically the local anomalies are detectable via infinitesimal gauge or diffeomorphism transformations that can be continuously deformed from the identity.}
\item all \emph{global anomalies} (\emph{nonperturbative anomalies}): \\
classified by finite abelian groups as a product of $\mathbb{Z}_n$ (the torsion classes) for some positive integer $n$. \\ 
e.g.,  Witten SU(2) \cite{Witten1982fp} and the new SU(2) anomalies \cite{WangWenWitten2018qoy1810.00844},
global gravitational anomalies \cite{Witten1985xe}.
{Typically the global anomalies are detectable only via large gauge or diffeomorphism transformations that \emph{cannot} be continuously deformed from the identity.}
\end{itemize}
More examples of $d$d anomalies characterized by $(d+1)$d cobordism invariants can be found in \cite{WanWang2018bns1812.11967,Witten2019bou1909.08775}.
} 
we propose an Ultra Unification that a new Topological Force 
comes into a theme of unification joining with three known fundamental forces and other hypothetical GUT forces. (See Figure \ref{fig:BSM-UU-tree}.)
More concretely, there is a new gapped Topological Phase Sector
whose underlying dynamical gauge interactions are the Topological Forces.
Alternatively, there could also be ``right-handed sterile'' neutrinos, gapless or more general interacting conformal field theories, or their combinations 
 to altogether cancel the mixed gauge-gravitational anomaly (enumerated in \Sec{sec:Consequences}).
In a modern perspective, we should view the SM and GUT all as {\bf effective field theories (EFT)} suitable below certain energy scales.
Whenever ``elementary particles'' are mentioned, they only mean to be ``elementary field quanta with respect to a given EFT.''
Likewise, Ultra Unification should be viewed as an EFT 
which contains SM or GUT but also additional gapped Topological Phase Sectors with 
low energy Lorentz invariant unitary topological quantum field theories (TQFTs)
of Schwarz type (which is the 4d analog of the 3d Chern-Simons-Witten theories \cite{ChernSimons1974ft, Schwarz1978cn, Witten1988hfJonesQFT}),
or additional neutrinos, or additional gapless or conformal sectors.
Topological Force and the gapped Topological Phase Sector here have specific physical and mathematical meanings, which we will clarify in
\Sec{sec:TopologicalSectorForce}.
Before digging into Topological Phase Sector, we should state the assumptions and the logic that lead to the assertion of 
Ultra Unification, in \Sec{sec:Logic}.\footnote{{\bf Conventions:} We follow the conventions of \Refe{WW2019fxh1910.14668, JW2006.16996,JW2008.06499}.
We denote $n$d for $n$-dimensional spacetime. We also follow the modern condensed matter or extreme quantum matter terminology on
the interacting phases of quantum matter \cite{Wen2016ddy1610.03911, Senthil1405.4015}. For example, \\
$\bullet$ \emph{Long-range entangled gapped} topological phases, whose low energy
describes the noninvertible TQFTs, are known as intrinsic topological orders, which include examples of fractional quantum Hall states.\\
$\bullet$ \emph{Short-range entangled gapped} topological phases protected by some global symmetry $G$, whose low energy
describes the invertible TQFTs, are known as symmetry-protected topological states (SPTs) \cite{Chen2011pg1106.4772}, which include examples of topological insulators and topological superconductors \cite{2010RMP_HasanKane, 2011_RMP_Qi_Zhang}.\\
$\bullet$ By a noninvertible TQFT, it means that the absolute value partition function $|{\bf Z}(M)| \neq 1$ on a generic spacetime manifold $M$ with nontrivial topology (e.g., cycles or homology classes).\\
$\bullet$ By an invertible TQFT (iTQFT), 
it means that the absolute value partition function $|{\bf Z}(M)| = 1$ on any spacetime manifold $M$ with any topology. Thus ${\bf Z}(M)$ implies the existence of an inverted phase
${\bf Z}'(M)\equiv {\bf Z}(M)^{-1}$ which defines another iTQFT ${\bf Z}'(M)$ that can cancel with the original iTQFT ${\bf Z}(M)$, as the stacking of two iTQFTs become a trivial vacuum
${\bf Z}(M) \cdot {\bf Z}'(M)=1$ for any $M$.\\
$\bullet$ By a trivial gapped vacuum with no TQFT or trivial TQFT, it means
that the partition function ${\bf Z}(M) = 1$ on any spacetime manifold $M$ with any topology.\\
\label{ft:Conventions}
We absolutely should distinguish the above beyond-Ginzburg-Landau quantum phases 
(long-range entangled and short-range entangled states)
from the within-Ginzburg-Landau symmetry-breaking phases
(long-range and short-range orders and correlations).
See more in \Sec{label:Summary}.
}

\section{Logic to Ultra Unification}
\label{sec:Logic}

\subsection{Assumptions}
\label{sec:Assumptions}
Our logic \cred{leading} to Ultra Unification starts with the three {\bf Assumptions} mostly given by Nature and broadly confirmed by experiments:
%
\begin{enumerate}
\item
 {\bf Standard Model gauge group} $G_{\text{SM}_q}$:
 The Standard Model gauge theory has a local Lie algebra $su(3) \times su(2) \times u(1)$,
but the global structure of Lie group $G_{\text{SM}_q}$ has four versions:\footnote{\cred{We denote 
the lower-case $su, so, \dots$ for the Lie algebra and the upper-case $\SU, \SO, \dots$ for the Lie group.}}
\bea
G_{\text{SM}_q} \equiv \frac{\SU(3) \times   \SU(2) \times \U(1)}{\Z_q},  \quad \text{ with } q=1,2,3,6.
\eea
All the quantum numbers of quarks and leptons are compatible with the representations of any version of $q=1,2,3,6$.
To confirm which version is used by Nature, it requires the experimental tests on extended objects such as 1d line or 2d surface operators (see recent expositions in \cite{AharonyASY2013hdaSeiberg1305.0318, Tong2017oea1705.01853, Ang2019txyAngRoumpedakis,Wan2019sooWWZHAHSII1912.13504}).

{\bf SU(5) and Spin(10) gauge group}: Conventionally, people write the 
Georgi-Glashow model
\cite{Georgi1974syUnityofAllElementaryParticleForces} as the $su(5)$ GUT  
and Fritzsch-Minkowski model \cite{Fritzsch1974nnMinkowskiUnifiedInteractionsofLeptonsandHadrons} as
the $so(10)$ GUT because they have the local Lie algebra $su(5)$ and $so(10)$, respectively.
However, they have the precise global Lie group SU(5) and Spin(10), respectively.
Only $q=6$, we are allowed to have the embedding of SM gauge group 
$G_{\text{SM}_6} \equiv \frac{\SU(3) \times   \SU(2) \times \U(1)}{\Z_6} \subset \SU(5)$ into the $su(5)$ GUT (see more discussions later in \eq{eq:embedding}).

\item  {\bf  Observed 15 Weyl fermions per generation by experiments}:
So far the HEP experiments only confirmed the 15 Weyl fermions per generation of SM.\footnote{Here 
Weyl fermions are spacetime Weyl spinors, which is 
${\bf 2}_L$ {of} ${{\Spin(1,3)}}=\rm{SL}(2,\C)$  with a complex representation in the Lorentz signature.
On the other hand,
the Weyl spinor is 
${\bf 2}_L$  {of} $\Spin(4)=\SU(2)_L \times \SU(2)_R$ with a {pseudoreal representation}
in the Euclidean signature.}
The experimentalists have \emph{not} yet confirmed the existence of 16th Weyl fermion for each generation.
The 16th Weyl fermion (or possibly even more than 16 fermions per generation) 
is also known as the right-handed neutrino $\nu_R$ or the sterile neutrino.
Namely, given the number of generation (or family)
$N_{\text{gen}}=3$,
we have {$15 \times N_{\text{gen}}= 45$} Weyl fermions confirmed in the SM.
Similarly, we consider   {$15 \times N_{\text{gen}}= 45$} Weyl fermions applicable to 
the $su(5)$ GUT.\footnote{However, the $so(10)$ GUT requires 16 Weyl fermions per generation due to the fermions sit at the {\bf 16} of the Spin(10).
This fact is used to argue the possibility of topological quantum phase transition between the energy scale of 
the 15n Weyl-fermion $su(5)$ GUT and 16n Weyl-fermion $so(10)$ GUT in \Refe{JW2008.06499}.}
(The quantum numbers and representations of the elementary particles in SM and GUT can be found in Table 1 and 2 of \cite{JW2006.16996}.)
{In terms of the quantum numbers of the 15 Weyl fermions per generation 
of $su(3) \times su(2) \times u(1)$ SM and that of $su(5)$ GUT {as chiral gauge theories}, {in the left-handed ($L$) Weyl spinor basis,} they are in the representations:
\begin{eqnarray}
(\overline{\bf 3},{\bf 1},1/3)_{L} \oplus ({\bf 1},{\bf 2},-1/2)_L 
\oplus
 ({\bf 3},{\bf 2}, 1/6)_L \oplus (\overline{\bf 3},{\bf 1}, -2/3)_L \oplus ({\bf 1},{\bf 1},1)_L \text{ of $su(3) \times su(2) \times u(1)$}\cr
\sim  \overline{\bf 5} \oplus {\bf 10} \text{ of $su(5)$.}\nonumber
\end{eqnarray}
Adding the 16th Weyl fermion (the sterile neutrino) as ${({\bf 1},{\bf 1},0)_L}$ of  $su(3) \times su(2) \times u(1)$ gives us
$\overline{\bf 5} \oplus {\bf 10} \oplus {\bf 1}$ of $su(5)$ also  ${\bf 16}^+$ of $so(10)$ (precisely the 16-dimensional spinor representation of $\Spin(10)$).
}

\item {\bf A variant discrete Baryon minus Lepton number (${ \mathbf{B}-  \mathbf{L}}$) {is preserved} at high energy}:
We hypothesize 
a discrete ${X}$ symmetry,
which is a modified version of $({ \mathbf{B}-  \mathbf{L}})$ number up to some electroweak hypercharge $Y$ \cite{Wilczek1979hcZee} is preserved
(preserved at least at a higher energy):\footnote{Follow \cite{WW2019fxh1910.14668, JW2006.16996, JW2008.06499}, 
we choose the convention 
that the $\U(1)_{\rm{EM}}$ electromagnetic charge is
$Q_{\rm{EM}}=T_3 +Y$. 
The $\U(1)_{\rm{EM}}$ is the unbroken (not Higgsed) electromagnetic gauge symmetry
and $T_3= \frac{1}{2}
\begin{pmatrix}
1 & 0\\
0 & -1
\end{pmatrix}$ is a generator of SU(2)$_{\text{weak}}$.
The electroweak Higgs field is in the representation
${({\bf 1},{\bf 2}, \frac{1}{2})}$ of  $su(3) \times su(2) \times u(1)$.
}
\bea
X &\equiv&
5({ \mathbf{B}-  \mathbf{L}})-4Y.
\eea
The importance of this discrete symmetry $\Z_{4,X}$ (as a mod 4 symmetry of $\U(1)_{X}$), in the context of global anomalies for SM and GUT
is emphasized by Garcia-Etxebarria-Montero \cite{GarciaEtxebarriaMontero2018ajm1808.00009}.

It is easy to check that (e.g., see the Table 1 and 2 of \cite{JW2006.16996}):\\ 
$\bullet$ all the particles from $\bar{\bf 5}$ of $su(5)$ GUT has a $\U(1)_{X}$ charge $-3$.\\
$\bullet$ all the particles from {\bf 10} of $su(5)$ GUT has a $\U(1)_{X}$ charge $+1$.\\
$\bullet$ the singlet right-handed neutrino (if any) is in {\bf 1} of $su(5)$ GUT with a $\U(1)_{X}$ charge $+5$.\\ 
$\bullet$ all the fermions of SM has a $\Z_{4,X}$ charge $+1$. \\ 
$\bullet$ the electroweak Higgs $\phi$ has a $\U(1)_{X}$ charge $-2$, thus a $\Z_{4,X}$ charge $+2$.\\ 
The $\Z_{4,X}$ also contains the fermion parity $\Z_2^F$ (whose operator $(-1)^F$ gives $(-1)$ to all fermions) as a normal subgroup
(so $\Z_{4,X}$ generator square $X^2=(-1)^F$ as the fermion parity):
\bea
\U(1)_{X} \supset \Z_{4,X} \supset \Z_2^F.
\eea
By looking at these consistent $\Z_{4,X}$ quantum number of SM particles, it is natural to hypothesize 
the discrete $X$ symmetry plays an important role at a higher energy above the SM energy scale.
\end{enumerate}
In \Sec{sec:Anomaly}, we review the 
anomaly and cobordism constraints given in \cite{WW2019fxh1910.14668, JW2006.16996, JW2008.06499}.
Readers can freely skip the technical discussions on anomalies, and directly
go to the final logic step lead to Ultra Unification in \Sec{sec:Consequences}.
\subsection{Anomaly and Cobordism Constraints}

\label{sec:Anomaly}

Based on the three mild and widely accepted assumptions listed in \Sec{sec:Assumptions},
we then impose the constraints from all invertible quantum anomalies via the cobordism calculation on SM and GUT models.
The purpose is to check the consistency of the 15n Weyl fermion SM and GUT models:
$$
\text{{\bf Check:}  Perturbative local and  nonperturbative global anomalies classified via cobordism.}
$$
The classification of $d$d 't Hooft anomalies of global symmetries $G$ is equivalent to the 
classification of $(d+1)$d invertible TQFTs with $G$-symmetry defined on a $G$-structure manifold,\footnote{For the QFT setup, 
we only require the category of smooth, differentiable, and triangulable manifolds.}
given 
by the cobordism group data
$\Omega^{d}_{G} \equiv \TP_d(G)$ defined in Freed-Hopkins 
\cite{Freed2016}.\footnote{Let us compare 
the cobordism group
$\Omega^{d}_{G} \equiv \TP_d(G)$ defined in Freed-Hopkins \cite{Freed2016}
and the more familiar bordism group
$\Omega_{d}^{G}$.
Here the cobordism group
$\Omega^{d}_{G} \equiv \TP_d(G)$
 \emph{not only} contains $\mathrm{Hom}(\Omega^{G,\mathrm{tors}}_{d}, \mathrm{U(1)})$ 
(the Pontryagin dual of the torsion subgroup (= tors) of the
bordism group $\Omega_{d}^{G}$), \emph{but also} contains the integer $\mathbb{Z}$ classes (the free part) 
descended from the free part of the bordism group $\Omega_{d+1}^{G,\mathrm{free}}$ of one higher dimension.
In other words, \\
$\bullet$ The classification of $(d-1)$d \emph{nonperturbative global anomalies}
can be read from the torsion part (the finite subgroup part) of
cobordism group $\Omega^{d,{\mathrm{tors}}}_{G} \equiv \TP^{\mathrm{tors}}_{d}(G)$. It can also be read from the  torsion part of the bordism group $\Omega^{G,\mathrm{tors}}_{d}$ data.\\
$\bullet$ The classification of $(d-1)$d \emph{perturbative local anomalies}
can be read from the free part (the $\mathbb{Z}$ classes) of
cobordism group $\Omega^{d,\mathrm{free}}_{G} \equiv \TP^{\mathrm{free}}_{d}(G)$, 
also from the free part of the bordism group $\Omega^{G,\mathrm{free}}_{d+1}$ data.\\
In this work, we concern the most for $(d-1)=4$ and $d=5$.}
The symmetry 
\bea
G\equiv ({\frac{{G_{\text{spacetime} }} \ltimes  {{G}_{\text{internal}} } }{{N_{\text{shared}}}}}) \equiv {{G_{\text{spacetime} }} \ltimes_{{N_{\text{shared}}}}  {{G}_{\text{internal}} } }
\eea
contains the spacetime symmetry
${G_{\text{spacetime} }}$ 
and
the internal symmetry ${{G}_{\text{internal}} }$.\footnote{$\bullet$ The 
${G_{\text{spacetime} }}$ is the spacetime symmetry, such as the spacetime rotational symmetry
$\SO\equiv \SO(d)$
or the fermionic graded spacetime rotational Spin group symmetry $\Spin\equiv \Spin(d)$.\\
$\bullet$  The ${{G}_{\text{internal}} }$ is the internal symmetry, such as
${{G}_{\text{internal}} }$ in the SM as
$G_{\text{SM}_q} \equiv \frac{\SU(3)\times \SU(2)\times \U(1)}{\Z_q}$ with $q=1,2,3,6.$
We also have ${{G}_{\text{internal}} }=\SU(5)$ in the $su(5)$ 
GUT and ${{G}_{\text{internal}} }=\Spin(10)$ in the $so(10)$ or Spin(10) GUT.
The ${N_{\text{shared}}}$ is the shared common normal subgroup symmetry between ${G_{\text{spacetime} }}$ 
and ${{G}_{\text{internal}} }$.
{The  ``semi-direct product $\ltimes$''   
extension is due to a group extension from ${{G}_{\text{internal}} }$ by ${G_{\text{spacetime} }}$. For a trivial extension,
the semi-direct ``$\ltimes$'' becomes a direct product ``$\times$.'' }}
Our perspective is that:\\ 
$\bullet$ We can treat the spacetime-internal $G$ as a global symmetry,
and we view the anomaly associated with $G$ as 't Hooft anomalies 
\cite{tHooft1979ratanomaly}
of $G$ symmetry.\\[2mm]
$\bullet$ Then, we can ask all obstructions to dynamically gauging the ${{G}_{\text{internal}} }$ as a gauge group,
which give rise to all the dynamical gauge anomaly cancellation conditions that any consistent gauge theory must obey.\\[2mm]
To proceed, we follow the results of \cite{WW2019fxh1910.14668, JW2006.16996, WanWangv2},
the relevant total spacetime-internal symmetry $G$ for the SM$_q$
is 
$G=\Spin(d) \times_{\Z_2^F} \Z_{4,X}\times G_{\SM_q}$ with $q=1,2,3,6$,
and for the $su(5)$ GUT is
$G= \Spin(d) \times_{\Z_2^F} \Z_{4,X}\times \SU(5)$.
Only the $q=6$ case of $\SM_6$ can be embedded into the $su(5)$ GUT.

\subsubsection{Standard Models}

\Refe{WW2019fxh1910.14668} considers
the classification of
$G$-anomalies in 4d given by the $d$=5 cobordism group 
$\Omega^{d}_{G} \equiv \TP_d(G)$ for all {\bf Standard Models} of $\SM_q$ with an extra discrete $\Z_{4,X}$ symmetry:
\bea
\TP_{d=5}(\Spin \times_{\Z_2^F} \Z_{4,X} \times G_{\SM_q})=
\left\{ \begin{array}{ll} 
\Z^5\times\Z_2\times\Z_4^2\times\Z_{16}, & \quad q=1,3.\\
 \Z^5\times\Z_2^2\times\Z_4\times\Z_{16}, &\quad q=2,6.
 \end{array}
\right.
\eea
Here we summarize the anomaly classification for the Standard Model 
$G_{\SM_q} \equiv \frac{\SU(3) \times   \SU(2) \times \U(1)_Y}{\Z_q}$
obtained in \cite{WW2019fxh1910.14668, JW2006.16996, WanWangv2}.
Below we write the 4d anomalies in terms of the 
5d cobordism invariants or invertible topological quantum field theories (iTQFTs). 
{For perturbative local 4d anomalies, we can also write them customarily as the 6d anomaly polynomials, and their cubic terms of gauge or gravitational couplings in the 
one-loop triangle Feynman diagram. Here is the list of classifications of anomalies}:\footnote{Here
we 
follow the conventions of \cite{WanWang2018bns1812.11967, WW2019fxh1910.14668, JW2006.16996, JW2008.06499}:\\
$\bullet$ We can characterize anomalies via (perturbative) \emph{local} anomalies or (nonperturbative) \emph{global} anomalies.\\
$\bullet$ We can also characterize anomalies via their induced fields: 
\emph{pure gauge} {anomalies}, \emph{mixed gauge-gravity} {anomalies}, or \emph{gravitational} anomalies {(those violate the general covariance under coordinate reparametrization; i.e. diffeomorphism)}.\\
 $\bullet$ The $c_j(G)$ is the $j$th Chern class of the associated vector bundle of the principal $G$-bundle.\\
$\bullet$  We will use $\text{CS}_{2n-1}^V$ to denote the Chern-Simons $(2n-1)$-form 
for the Chern class (if $V$ is a complex vector bundle) 
or the Pontryagin class (if $V$ is a real vector bundle).  
The relation between the Chern-Simons form and the Chern class is
$
c_n(V)=\dd \text{CS}_{2n-1}^V
$ 
where the $\dd$ is the exterior differential and the $c_n(V)$ is regarded as a closed differential form in de Rham cohomology.
The $w_j(TM)$ is the $j$-th Stiefel-Whitney class of spacetime tangent bundle $TM$ of the base manifold $M$.\\
$\bullet$  The PD is defined as the Poincar\'e dual.
We define the product notation $\rc \upeta$ between a cohomology class $\rc$ and a fermionic invariant $\upeta$ via the Poincar\'e dual 
PD of cohomology class, thus $\rc \upeta \equiv \eta (\PD(\rc))$.
We use the $\smile$ notation for the cup product between cohomology classes.
We often make the cup product $\smile$ and the Poincar\'e dual PD implicit.
\\
$\bullet$  {The $\mu$ is the 3d Rokhlin invariant.
If $\partial M^4=M^3$, then $\mu(M^3)=(\frac{\sigma-\rF\cdot\rF}{8} )(M^4)$, thus
$\mu(\text{PD}(c_1(\U(1))))$ is related to $\frac{1}{8}(\sigma-\rF \cdot \rF )(\PD(c_1(\U(1))))$.
Here $\cdot$ is the intersection form of $M^4$.
The $\rF $ is the characteristic 2-surface in a 4-manifold $M^4$, it obeys the condition $\rF\cdot x=x\cdot x\mod2$ for all $x\in\H_2(M^4,\Z)$. 
By the Freedman-Kirby theorem: $(\frac{\sigma-\rF\cdot\rF}{8} )(M^4)=\text{Arf}(M^4,\rF)\mod2$,
 here $\text{Arf}(M^4, \rF)$ is defined to be Arf(q) where q: $\H_1(\rF,\Z_2) \to \Z_2$ 
 is a quadratic form associated with the characteristic surface $F \subset M^4$ \cite{Saveliev}.}\\
$\bullet$ The $\tilde{\eta}$ is a mod 2 index of 1d Dirac operator as a cobordism invariant of the bordism group $\Omega_1^{\Spin}=\Z_2$.\\ 
$\bullet$ The ${\eta}'$ is a mod 4 index of 1d Dirac operator as a cobordism invariant of the bordism group {$\Omega_1^{\Spin\times_{\Z_2} \Z_4}=\Z_4$}.\\ 
$\bullet$ The Arf invariant \cite{Arf1941} is a 2d cobordism invariant of $\Omega_2^{\Spin}=\Z_2$.
The Arf appears to be the low energy iTQFT of a 1+1d Kitaev fermionic chain \cite{Kitaev2001chain0010440},
whose boundary hosts a single 0+1d real Majorana zero mode on each of open ends.\\
$\bullet$ {We use the notation ``$\sim$'' to indicate the two sides are equal in that dimension up to a total derivative term.}\\
$\bullet$ {Because of the $\Z_{4,X} \supset \Z_2^F$, we have a short exact sequence 
$
0 \to \Z_2^F \to \Z_{4,X}  \to \frac{\Z_{4,X}}{\Z_2^F} \to 0.
$ 
Together with the SO structure,
\bea \label{eq:Spin-Z4}
1 \to \Z_2^F \to  \Spin \times_{{\Z_2^F}} \Z_{4,X}  \to \SO \times \frac{\Z_{4,X}}{\Z_2^F} \to 1.
\eea
Below we use the standard convention, 
all cohomology classes are pulled back to the manifold $M$ along the maps, 
thus the gauge field in $\H^1(\B {\rm A}, {\rm A})$ can be pulled back to the gauge field in $\H^1(M, {\rm A})$ for some abelian group ${\rm A}$.
\Eqn{eq:Spin-Z4} is similar to 
$1 \to \Z_2^F \to  \Spin^c \equiv \Spin \times_{{\Z_2^F}} \U(1)  \to \SO \times \frac{\U(1)}{\Z_2^F} \to 1$,
so below we compare $\Spin^c$ with $\Spin \times_{{\Z_2^F}} \Z_{4,X}$ gauge fields.
\\
$\bullet$
The $\Spin^c$ gauge field $A$ is \emph{not} an ordinary abelian gauge field, 
due to $\frac{1}{2}w_2(TM) = \frac{\dd A}{2 \pi}  \mod \Z$ thus $w_2(TM) = c_1 =\frac{\dd (2A)}{2 \pi} \mod 2$ \cite{SeibergWitten1602.04251};
but the $2A$ is an ordinary abelian gauge field in $\H^1(M, \frac{\U(1)}{\Z_2^F})$.\\
$\bullet$ 
The $\Spin \times_{{\Z_2^F}} \Z_{4,X}$ gauge field $\CA_{\Spin \times_{{\Z_2^F}} \Z_{4,X}}$ denoted as 
$\CA_{{\Z_4}}$ in brief is \emph{not} an ordinary abelian gauge field.
We define the cohomology classes of background gauge field
 $\CA_{{\Z_2}} \in \H^1(M,\frac{\Z_{4,X}}{\Z_2^F})$,
which on a ${\Spin \times_{\Z_2^F} {\Z_{4,X}}}$-manifold $M$ obeys a constraint:
$w_2(TM) = \CA_{{\Z_2}}^2 = \Sq^1(\CA_{{\Z_2}}) = \frac{1}{2} \delta \CA_{{\Z_2}}$ 
with a coboundary operator $\delta$ ,
where $\Sq^1(\CA_{{\Z_2}})$ 
 is defined as the Steenrod square map $\H^1(M,\frac{\Z_{4,X}}{\Z_2^F}) \overset{\Sq^1}{\longrightarrow} \H^2(M, {\Z_2^F})$.
If we can lift the $\CA_{{\Z_2}}$ gauge field to $p^*(\CA_{{\Z_2}})$  via a pullback
$\Spin \times {\Z_{4,X}} \overset{p}{\longrightarrow} {\Spin \times_{\Z_2^F} {\Z_{4,X}}}$, this requires $w_2(TM) = \CA_{{\Z_2}}^2 =0$.
\emph{Only if} this pullback exists, then $\CA_{{\Z_4}}=p^*(\CA_{{\Z_2}})$ becomes an ordinary abelian ${\Z_4}$-gauge field such that 
$\CA_{{\Z_2}} = \CA_{{\Z_4}} \mod 2$,
so $\CA_{{\Z_2}}^2 = (\CA_{{\Z_4}} \mod 2)^2 =0$ because 
$\frac{1}{2} \delta \CA_{{\Z_2}} =\frac{1}{2} \delta( \CA_{{\Z_4}} \mod 2)= 2 (\frac{1}{4} \delta \CA_{{\Z_4}})=0 \mod 2$.
}
\label{ft:convention}
}
\begin{enumerate}[leftmargin=-0mm]
\item 
{U(1)$_Y^3$: 4d $\Z$ class local pure gauge
anomaly  from 5d $\text{CS}_1^{\U(1)}c_1(\U(1))^2$ and 6d $c_1(\U(1))^3$}.
\item
{U(1)$_Y$-SU(2)$^2$: 4d $\Z$ class local pure gauge
anomaly  from 5d $\text{CS}_1^{\U(1)}c_2(\SU(2))$ and 6d $c_1(\U(1))c_2(\SU(2))$}.
\item {U(1)$_Y$-SU(3)$^2$: 4d $\Z$ class local pure gauge anomaly from 5d $\text{CS}_1^{\U(1)}c_2(\SU(3))$ and 6d $c_1(\U(1))c_2(\SU(3))$}.
\item {U(1)$_Y$-(gravity)$^2$: 4d $\Z$ class local mixed gauge-gravity anomaly from 5d $\mu(\text{PD}(c_1(\U(1))))$ and 6d 
$\frac{c_1(\U(1))(\sigma-\rF \cdot \rF )}{8} \equiv \frac{1}{8}(\sigma-\rF \cdot \rF )(\PD(c_1(\U(1))))$}.
\item {SU(3)$^3$: 4d $\Z$ class local pure gauge anomaly from 5d $\frac{1}{2}{\text{CS}_5^{\SU(3)}}$ and 6d $\frac{1}{2}{c_3(\SU(3))}$}.
\item {Witten SU(2) anomaly}: 4d $\Z_2$ class global mixed gauge-gravity anomaly from 5d $\tilde\eta\,c_2(\SU(2)) \equiv \tilde\eta(\PD(c_2(\SU(2))))$ and 
6d $\text{Arf}\,c_2(\SU(2))\equiv \text{Arf}(\PD(c_2(\SU(2))))$.
However, the original Witten $\Z_2$ global anomaly becomes \emph{mutated}: \\
$\bullet$ When $q=1$ or $3$, Witten anomaly \emph{mutated} to a 4d $\Z_4$ class global mixed gauge-gravity anomaly, given by a 5d cobordism invariant
$\eta'\,c_2(\SU(2))\equiv \eta'(\PD(c_2(\SU(2))))$.
There is a short exact sequence $0 \to \Z_2 \to \Z_4 \to \Z_2 \to 0$ where 
the $(\CA_{{\Z_2}}) c_2(\SU(2))$ sits at the $\Z_2$ normal subgroup
and 
the original Witten anomaly $\tilde\eta\,c_2(\SU(2))$ sits at the $\Z_2$ quotient group,
while the mutated $\eta'\,c_2(\SU(2))$ sits at the $\Z_4$ total group. 
{If we write the group element $k =  2 k_N + k_Q \in \Z_4$ via two data $(k_N, k_Q)$ from the normal and quotient groups,
then
$k \eta' = ( 2 k_N + k_Q) \eta' =  k_N \CA_{{\Z_2}} + k_Q \tilde \eta $, 
also $k \eta'\,c_2(\SU(2))= ( 2 k_N + k_Q) \eta'\,c_2(\SU(2))
=k_N \CA_{{\Z_2}} \, c_2(\SU(2)) + k_Q \tilde\eta \, c_2(\SU(2))$.}
\\
$\bullet$ When $q=2$ or $6$, Witten anomaly \emph{mutated} to become part of 4d $\Z$ class local mixed gauge-gravity anomaly (firstly explained in \cite{Davighi2020bvi2001.07731}),
given by a 5d cobordism invariant 
{$\frac{1}{2}\text{CS}_1^{\U(2)}c_2(\U(2)) \sim
\frac{1}{2}c_1(\U(2))\text{CS}_3^{\U(2)}$}.
There is 
a short exact sequence $0 \to \Z \overset{2}{\to} \Z \to \Z_2 \to 0$
where 
the {$\text{CS}_1^{\U(1)}c_2(\SU(2)) \sim c_1(\U(1))\text{CS}_3^{\SU(2)}$}
sits at the $\Z$ normal subgroup 
and 
the original Witten anomaly $\tilde\eta \, c_2(\SU(2))$ 
sits at the $\Z_2$ quotient group,
while the mutated {$\frac{1}{2}\text{CS}_1^{\U(2)}c_2(\U(2))
\sim
\frac{1}{2}c_1(\U(2))\text{CS}_3^{\U(2)}$} sits at the $\Z$ total group. 
{If we write the group element $k =2 \frac{k_{\rm even}}{2} + (k \mod 2) \in \Z$ via two data $(\frac{k_{\rm even}}{2}, k \mod 2)$ 
from the normal and quotient groups,
then
$k  \frac{1}{2}\text{CS}_1^{\U(2)}c_2(\U(2))
= \frac{k_{\rm even}}{2} \text{CS}_1^{\U(1)}c_2(\SU(2)) 
+ (k \mod 2) \tilde\eta \,  c_2(\SU(2))$.
}
\item $(\CA_{{\Z_2}}) c_2(\SU(2))$: 4d $\Z_{2}$ global gauge anomaly given by a 5d cobordism invariant 
$\CA_{{\Z_2}} \smile c_2(\SU(2))$.
\\
$\bullet$ When $q=1$ or $3$, as explained earlier, it fuses with Witten anomaly to become a $\Z_4$ global anomaly. \\
$\bullet$ When $q=2$ or $6$, this $\Z_{2}$ global anomaly occurs.
\item {$(\CA_{{\Z_2}}) c_2(\SU(3))$: 4d $\Z_{2}$ global gauge anomaly.}\\ 
$\bullet$ When $q=1$ or $2$, this is given by a 5d cobordism invariant 
$\CA_{{\Z_2}} \smile c_2(\SU(3))$.
\\
$\bullet$ When $q=3$ or $6$, this is given by a 5d cobordism invariant 
$\CA_{{\Z_2}} \smile c_2(\U(3))$.
%
\item {$\eta'\, c_1(\U(1))^2$: 4d $\Z_{4}$ global mixed gauge-gravity anomaly.}\\ 
$\bullet$ {When $q=1$, this is given by a 5d cobordism invariant $\eta'\, c_1(\U(1))^2 \equiv \eta'(\PD(c_1(\U(1))^2) )$.}\\
$\bullet$ {When $q=2$, this is given by a 5d cobordism invariant $\eta'\, c_1(\U(2))^2 \equiv \eta'(\PD(c_1(\U(2))^2) )$.}\\
$\bullet$ {When $q=3$, this is given by a 5d cobordism invariant $\eta'\, c_1(\U(3))^2 \equiv \eta'(\PD(c_1(\U(3))^2) )$.}\\
$\bullet$ {When $q=6$, this is given by a 5d cobordism invariant $\eta'\, c_1(\U(2))^2 \sim \eta'\,  c_1(\U(3))^2$.}
\item {$\eta(\text{PD}(\CA_{{\Z_2}}))$:  4d $\Z_{16}$ global mixed gauge-gravity anomaly is given by a 5d cobordism invariant {$\eta(\text{PD}(\CA_{{\Z_2}}))$}. }
The $\eta(\text{PD}(\CA_{{\Z_2}}))$ is
the value of Atiyah-Patodi-Singer (APS \cite{Atiyah1975jfAPS}) 
eta invariant ${\eta} \in \Z_{16}$ on the Poincar\'e dual (PD) submanifold of $\CA_{{\Z_2}}$.
It descends from the bordism group calculation of $\Omega_5^{\Spin \times_{\Z_2} \Z_4}=\Omega_4^{\Pin^+} =\Z_{16}$.
Many previous works had also explored these 5d $\Z_{16}$ fermionic invariants \cite{2018arXiv180502772T,Hsieh2018ifc1808.02881,GuoJW1812.11959}.
The relation between 5d $\eta(\text{PD}(\CA_{{\Z_2}}))$ and 4d $\eta$ is given by the Smith homomorphism \cite{Kapustin1406.7329, Hason1910.14039}.
\end{enumerate}
\subsubsection{Georgi-Glashow $su(5)$ Grand Unification}

\Refe{WW2019fxh1910.14668, WanWangv2} consider
the classification of
$G$-anomalies in 4d given by the $d$=5 cobordism group 
$\Omega^{d}_{G} \equiv \TP_d(G)$ for  {\bf the $su(5)$ GUT} with an extra discrete $\Z_{4,X}$ symmetry:
\bea
\TP_{d=5}(\Spin \times_{\Z_2^F} \Z_{4,X} \times \SU(5))=
\Z \times\Z_2\times \Z_{16}.
\eea
Below we also write down the 4d anomalies in terms of the 
5d cobordism invariants or iTQFTs, or the 6d anomaly polynomials:
\begin{enumerate}[leftmargin=-0mm]
\item SU(5)$^3$: 4d $\Z$ class local gauge anomaly.
It is given by a 5d cobordism invariant $\frac{1}{2}\text{CS}_5^{\SU(5)}$, or more precisely 
{$\frac{1}{2}({(\CA_{{\Z_2}})^2\text{CS}_3^{\SU(5)}+\text{CS}_5^{\SU(5)}})$} including the $\Spin \times_{\Z_2^F} \Z_{4,X}$ contribution.
\item {$(\CA_{{\Z_2}}) c_2(\SU(5))$: 4d $\Z_{2}$ class global gauge anomaly}.
It is given by a 5d cobordism invariant {$(\CA_{{\Z_2}}) c_2(\SU(5))$} which detects a mixed anomaly between the 
$\frac{\Z_{4,X}}{\Z_2^F}$ gauge field and $\SU(5)$ gauge field.
\item {$\eta(\text{PD}(\CA_{{\Z_2}}))$:  4d $\Z_{16}$ global mixed gauge-gravity anomaly is given by a 5d cobordism invariant {$\eta(\text{PD}(\CA_{{\Z_2}} ))$}
which detects a mixed anomaly between the 
$\frac{\Z_{4,X}}{\Z_2^F}$ gauge field and the gravity.
This is again the same $\Z_{16}$ global anomaly from $\Omega_5^{\Spin \times_{\Z_2} \Z_4}=\Omega_4^{\Pin^+} =\Z_{16}$.}
\end{enumerate}

\subsection{Consequences lead to Ultra Unification}
\label{sec:Consequences}

\Refe{JW2006.16996, JW2008.06499} checked all the above local and global anomalies enlisted in \Sec{sec:Anomaly}
vanished for the SM$_q$ 
with $q=1,2,3,6$ and for the $su(5)$ GUT with 15 Weyl fermions per generation, 
except the 4d $\Z_{16}$ class global anomaly (the 
mixed gauge-gravitational anomaly probed by the discrete $\Z_{4,X}$ symmetry and fermionic spacetime rotational symmetry $\Spin(d)$
background fields) 
may \emph{not} be completely canceled.\footnote{We should briefly compare
the perspectives of Garcia-Etxebarria-Montero \cite{GarciaEtxebarriaMontero2018ajm1808.00009}, 
Davighi-Gripaios-Lohitsiri \cite{DavighiGripaiosLohitsiri2019rcd1910.11277}, and our previous result \cite{JW2006.16996, JW2008.06499}, 
and those with Wen \cite{WangWen2018cai1809.11171} or 
with Wan \cite{WW2019fxh1910.14668}. In terms of the relevancy to the 4d anomalies of SM and $su(5)$ GUT given the spacetime-internal symmetry $G$,\\
$\bullet$ Garcia-Etxebarria-Montero \cite{GarciaEtxebarriaMontero2018ajm1808.00009} checked: {$G=\Spin\times_{\Z_2}  \Z_4, \;  \Spin \times \SU(n),  \;  \Spin \times \Spin(n)$.}\\
$\bullet$ Wang-Wen \cite{WangWen2018cai1809.11171} checked: ${G=\frac{\Spin \times \Spin(n)}{{\Z_2^F}}, \frac{\Spin \times \Spin(10)}{{\Z_2^F}}, \;\Spin \times \SU(5)}$.\\
$\bullet$ Davighi-Gripaios-Lohitsiri \cite{DavighiGripaiosLohitsiri2019rcd1910.11277}  checked:
{$G=\Spin\times{G_{\text{SM}_{q}}}$, ${\Spin \times \Spin(n)}$, and other GUTs}.\\
$\bullet$ Wan-Wang \cite{WW2019fxh1910.14668, WanWangv2} checked: 
{$G=\Spin\times{G_{\text{SM}_{q}}}$, $\cred{\Spin\times_{\Z_2} \Z_4} \times {G_{\text{SM}_{q}}}$,}
{$\Spin \times \SU(5)$, $\Spin\times_{\Z_2}  \Z_4 \times  \SU(5)$,}
{${\Spin \times \Spin(n)}$, \; $\frac{\Spin \times \Spin(n)}{{\Z_2^F}}$ e.g., $n=10,18$, and other GUTs.}\\[2mm]
Thus, given by the starting assumption in \Sec{sec:Assumptions}, 
only Wan-Wang \cite{WW2019fxh1910.14668, WanWangv2} contains the complete anomaly classification data 
for $G=\Spin\times_{\Z_2} \Z_4 \times {G_{\text{SM}_{q}}}$ and $\Spin\times_{\Z_2}  \Z_4 \times  \SU(5)$
that we need for completing the argument. So we must have to employ the results of \Refe{WW2019fxh1910.14668}.
Although \Refe{GarciaEtxebarriaMontero2018ajm1808.00009,DavighiGripaiosLohitsiri2019rcd1910.11277} checked several global anomalies, 
but they do not exhaust checking all anomalies that we need for $G=\Spin\times_{\Z_2} \Z_4 \times {G_{\text{SM}_{q}}}$ and $\Spin\times_{\Z_2}  \Z_4 \times  \SU(5)$.\\[2mm]
Indeed, specifically for $G=\Spin\times_{\Z_2} \Z_4 \times {G_{\text{SM}_{q}}}$ and $\Spin\times_{\Z_2}  \Z_4 \times  \SU(5)$,
only \Refe{WW2019fxh1910.14668, WanWangv2} exhausted the cobordism classifications for all their possible anomalies,
and 
completed the {anomaly cancellation} 
checks on these groups.
 }
The cobordism invariant for any $\upnu \in \Z_{16}$ corresponds to a 5d iTQFT partition function  
\bea \label{eq:5dSPT}
{{\bf Z}_{{\text{$5$d-iTQFT}}}^{(\upnu)} }
\equiv
\exp(\frac{2\pi \ii}{16} \cdot\upnu \cdot \eta(\text{PD}(\CA_{{\Z_2}})) \bigg\rvert_{M^5}), & \text{ with } \eta \equiv \eta_{\text{Pin}^+} \in \Z_{16}, &\quad \upnu\in \Z_{16}.
\eea
Given a $G \supseteq {\Spin \times_{\Z_2^F} \Z_{4,X}}$ structure, the cohomology class 
$\CA_{{\Z_2}} \in \H^1(M,\Z_2)$ is the generator from  $\H^1(\B(\Z_{4,X}/\Z_2^F),\Z_2)$.
{To compute $\Omega_5^{\Spin  \times_{\Z_2^F} \Z_{4,X} \times \dots }=\Z_{16} \times \dots$, 
we use the Madsen-Tillmann ($MT$) spectra \cite{MadsenTillmann4} to obtain
$MT(\Spin  \times_{\Z_2^F} \Z_{4,X})=M\Spin \wedge \Sigma^{-2} M (\frac{\Z_{4,X}}{\Z_2^F})$ \cite{WW2019fxh1910.14668}, 
where $\wedge$ is the smash product 
and the $\Sigma$ denotes a suspension.\footnote{For a pointed topological space $\CX$, 
the $\Sigma$ denotes a suspension $\Sigma \CX=S^1\wedge \CX=(S^1\times \CX)/(S^1\vee \CX)$ 
where $\wedge$ and $\vee$ are smash product and wedge sum (a one point union) of pointed topological spaces respectively.}
Although 
the $\Spin \times_{{\Z_2^F}} \Z_{4,X}$ gauge field $\CA_{\Spin \times_{{\Z_2^F}} \Z_{4,X}}$ denoted as 
$\CA_{{\Z_4}}$ in brief is \emph{not} an ordinary abelian gauge field (see footnote \ref{ft:convention}),
the Thom spectra $M(\frac{\Z_{4,X}}{\Z_2^F})$ suggests that the cobordism invariant depends on the $\CA_{{\Z_2}}$ gauge field
(instead of the $\CA_{{\Z_4}}$ gauge field) in the cohomology class $\H^1(M,\frac{\Z_{4,X}}{\Z_2^F})$.}
The $\eta(\text{PD}(\CA_{{\Z_2}}))$ is
the value of APS eta invariant $\eta   \in \Z_{16}$ on the Poincar\'e dual (PD) submanifold of the cohomology class
$\CA_{{\Z_2}}$. This PD takes ${ \cap  \CA_{{\Z_2}}}$ (of the cohomology $\H^1$) from 5d to 4d (of the homology $\H_4$). 
The APS eta invariant $\eta \equiv \eta_{\text{Pin}^+}  \in \Z_{16}$ is the cobordism invariant of the bordism group  $\Omega_4^{\Pin^+} =\Z_{16}$.
The notation ``$\mid_{M^5}$'' means the evaluation on this invariant on a 5-manifold ${M^5}$.
{Note that $1 \to \Z_2^F \to \Pin^+ \to \O \to 1$ and $1 \to \Spin \to \Pin^+ \to \Z_2^T \to 1$ with the orthogonal group $\O$ and time-reversal symmetry $\Z_2^T$,
the Madsen-Tillmann ($MT$) spectra says $MT\Pin^+=M\Spin \wedge \Sigma^1MT\Z_2^T=M\Spin \wedge S^1 \wedge MT\Z_2^T$.}
We summarize the consequences and implications of this $\Z_{16}$ {anomaly non-vanishing for 15n Weyl fermions} in \Sec{sec:Consequences}.

\noindent
$\bullet$ \Refe{GarciaEtxebarriaMontero2018ajm1808.00009} used the existence of $\Z_{16}$ global anomaly and its anomaly {cancellation} 
to verify the conventional lore: the 16 Weyl fermions per generation scenario, by introducing a right-handed neutrino per generation. 

\noindent
$\bullet$ \Refe{JW2006.16996, JW2008.06499} take the $\Z_{16}$ anomaly as a secrete entrance to find hidden new sectors beyond the Standard Model, 
given the fact the HEP experiments only have detected 
 15 Weyl fermions per generation thus far.\\[-10mm] 

\paragraph{\fontsize{14}{16pt}\selectfont{Consequences}} 
Follow \Sec{sec:Assumptions} and \ref{sec:Anomaly}, the
$\Z_{16}$ anomaly index for both the SM$_q$ ($q=1,2,3,6$) and the $su(5)$ GUT
is the $(15 =-1  \mod 16)$ per generation, and {$(N_{\text{gen}}=3)$}.
Given the right-handed neutrino number $n_{\nu_{j,R}}$ (each ${\nu_{j,R}}$ has $\Z_{4,X}$ charge 1) 
for the $j$-th generation, and 
the new hidden sectors' anomaly index $\upnu_{\text{new sectors}}$,
we have the following anomaly {cancellation} 
condition to cancel the
$\Z_{16}$ anomaly:
\bea \label{eq:anomaly}
\boxed{
  (-{(N_{\text{gen}}=3)}+ 
(\sum_{j=e,\mu,\tau,\dots }n_{\nu_{j,R}} )
   + \upnu_{\text{new sectors}} ) = 0 \mod 16
.}
\eea
The question is: How to cancel the $\Z_{16}$ anomaly? We enlist as many \cred{Scenarios} as possible below.\\[-8mm]
\begin{enumerate}[leftmargin=-0mm, label=\textcolor{blue}{\arabic*}., ref={\arabic*}]
\item {\bf Standard Lore:} \label{1Standard}
We can introduce the right-handed neutrino (the 16th Weyl fermion) number $n_{\nu_{j,R}}=1$ for 
each generation (so $n_{\nu_{e,R}} = n_{\nu_{\mu,R}} = n_{\nu_{\tau,R}}=1$ for electron, muon, and tau neutrinos).
In this case, there is no {new hidden sector}.
 \begin{enumerate}[leftmargin=3mm, label=\textcolor{blue}{(1\alph*)}., ref={(1\alph*)}]
\item {\bf Massless}: \label{Massless}
$\Z_{4,X}$ can be preserved if fermions are gapless. 
\item {\bf Dirac mass}: \label{Dirac}
$\Z_{4,X}$ is also preserved by the Yukawa-Higgs-Dirac Lagrangian, but $\Z_{4,X}$ is spontaneously broken by the Higgs condensate
to give a {\bf Dirac} mass gap.
\item {\bf Majorana mass}: \label{Majorana}
$\Z_{4,X}$ is broken explicitly by {\bf Majorana} mass term.
\end{enumerate}

\item \label{2Proposals}
{\bf Proposals in \Refe{JW2006.16996, JW2008.06499}:} 
\Refe{JW2006.16996, JW2008.06499} proposed other novel ways to 
cancel the $\Z_{16}$ anomaly. Consequently, we can introduce new hidden sectors beyond the SM and the $su(5)$ GUT:
 \begin{enumerate}[leftmargin=3mm, label=\textcolor{blue}{(2\alph*)}., ref={(2\alph*)}]
\item \label{Z4XTQFT}
{$\Z_{4,X}$-symmetry-preserving anomalous gapped 4d topological quantum field theory (TQFT).}\footnote{In the context of 3d boundary and 4d bulk, 
a novel surface topological order was firstly pointed out by
Vishwanath-Senthil in an insightful work \cite{VishwanathSenthil1209.3058}. Later on many people follow up on developing the 
surface topological order constructions (see overviews in \cite{Senthil1405.4015, Wen2016ddy1610.03911} and \cite{Wang2017locWWW1705.06728}).
 \Refe{JW2006.16996} generalizes this condensed matter idea to find an anomalous symmetric 4d TQFT living on the boundary of 5d fermionic SPTs \Eq{eq:5dSPT}.
 Beware that although many 4d TQFTs listed below have n\"aive abelian gauge groups (like $[\Z_2]$ or $[\Z_4]$),
 but due to their fermionic nature and ${\Z_2^F}$-symmetry, these 4d TQFT can exhibit 
 nonabelian fusion rules or nonabelian braiding statistics, see \Sec{sec:TQFTPathIntegral} for discussions.}
We also call the finite \emph{energy gap} 
\bea \label{eq:DeltaTQFT}
\Delta_{\rm TQFT} \equiv E_{\text{excited}} - E_{\text{ground states}}
\eea 
for the first excitation(s) above the ground state sectors of this 4d TQFT 
as {\bf Topological mass} gap. The underlying quantum system has a 4d intrinsic topological order.
We name the anomaly index $\upnu_{\rm 4d}$ for this anomalous 4d TQFT.
\item \label{Z45dSPT}
{$\Z_{4,X}$-symmetry-preserving 5d invertible topological quantum field theory (iTQFT) 
given by the 5d cobordism invariant in \Eq{eq:5dSPT}.}
The underlying quantum system has a 5d symmetry-protected topological state (SPTs) with an \emph{extra bulk 5th dimension} whose
4d boundary can live the 4d Standard Model world.
We name the anomaly index $\upnu_{\rm 5d}$ to specify the boundary 4d anomaly of this 5d iTQFT.

\item \label{Z45dSET}
${\Z_2^F}$-symmetry-preseving
{5d bulk $[\frac{\Z_{4,X}}{\Z_2^F}]$-gauged TQFT and 4d boundary $[\Z_{4,X}]$-gauged TQFT}.
Overall the spacetime rotational symmetry is the fermionic Spin group $\Spin(d)$ graded the bosonic rotation special orthogonal group $\SO(d)$ by the fermion parity ${\Z_2^F}$.
\item \label{Z45dSET2}
{5d bulk $[{\Z_2^F}\times \frac{\Z_{4,X}}{\Z_2^F}]$-gauged TQFT and 4d boundary $[{\Z_2^F} \times \Z_{4,X}]$-gauged TQFT}.
Overall the spacetime rotational symmetry is the bosonic special orthogonal group $\SO(d)$.
{If the diffeomorphism symmetry of $\SO(d)$ or $\Spin(d)$ is further dynamically gauged, the outcome new sector may be a gravity theory or a topological gravity theory.}
\item \label{Z4XbreakTQFT}
$\Z_{4,X}$-symmetry-breaking gapped phase (e.g. Ginzburg-Landau paradigm phase or 4d TQFT).
\item \label{Z4XCFT}
$\Z_{4,X}$-symmetry-preserving or $\Z_{4,X}$-symmetry-breaking gapless phase, e.g., 
extra massless theories, free or interacting conformal field theories (CFTs).
{The interacting CFT with scale invariant gapless energy spectrum is also related to {\bf unparticle physics} \cite{Georgi200703260Unparticle} in
the high-energy phenomenology community.}
\end{enumerate}
\end{enumerate}
 
Scenarios \ref{Z4XTQFT} and \ref{Z45dSPT}, and their linear combinations, 
are the \emph{root phases} for new hidden gapped Topological Phase Sector.
For example, \emph{breaking} part of the global symmetries in
the linear combined \ref{Z4XTQFT} and \ref{Z45dSPT}
would give rise to Scenario \ref{Z4XbreakTQFT}.
For another example, \emph{gauging} part of the global symmetries in
the linear combined \ref{Z4XTQFT} and \ref{Z45dSPT}
would give rise to other Scenarios:\\
$\bullet$ Gauging $[\frac{\Z_{4,X}}{\Z_2^F}]$ in 5d and gauging $[\Z_{4,X}]$ in 4d gives
rise to Scenario \ref{Z45dSET}.\\
$\bullet$ Gauging $[{\Z_2^F}\times \frac{\Z_{4,X}}{\Z_2^F}]$ in 5d and gauging $[{\Z_2^F} \times \Z_{4,X}]$ in 4d gives
rise to Scenario \ref{Z45dSET2}.\\
The underlying quantum systems in Scenario \ref{Z4XTQFT} and \ref{Z45dSET}
have the symmetry-enriched topologically ordered state (SETs) in a condensed matter terminology.

We name the combination of the above anomaly {cancellation} 
scenarios,
including the standard lore (right-handed neutrinos in Scenario \ref{1Standard}) and the new proposals (all enlisted in Scenario \ref{2Proposals}) 
beyond the SM and the GUT, as the {\bf Ultra Unification}. 
Although introducing additional CFTs (\cred{Scenario \ref{Z4XCFT}}) to cancel the anomaly is equally fascinating,
we instead mostly focus on introducing the gapped Topological Phase Sector due to high-energy physics phenomenology (HEP-PH) constraints (see a summary in \cite{JW2006.16996}).
A central theme of Ultra Unification 
suggesting a new HEP  frontier
is that
\bea
\text{{\bf Ultra Unification}: HEP-PH provides {\bf Gapped Extended Objects}}\cr 
\quad \text{or {\bf Gapless Conformal Objects} {beyond Particle Physics}.}
\eea
Some more comments: \\
$\bullet$ These gapped extended objects (of 1d, 2d, 3d, $\dots$) 
are formulated mathematically in terms of \emph{gapped} TQFT extended operators (1d line, 2d surface, 3d brane, etc., topological operators), beyond the
0d particles.\\
$\bullet$ 
These extended operators are heavy in the sense that they sit at the energy scale above the TQFT energy gap  $\Delta_{\rm TQFT}$ (so
at or above the scale of $E_{\text{excited}}$ in \eq{eq:DeltaTQFT}).\\
$\bullet$  These extended operators are heavy in the sense that they have Topological mass and they can interact
with dynamical gravity. So these gapped extended objects may be the Dark Matter candidate. 
If the $\Delta_{\rm TQFT}$ is large, then the gapped extended objects are {\bf heavy Dark Matter} candidates, in contrast the
gapless conformal objects are {\bf light Dark Matter} candidates.\\
$\bullet$ There are fractionalized anyonic excitations at the open ends of topological operators (1d line, 2d surface, 3d brane, etc.).
In other words, the particle 1d worldline is the 1d line topological operator.
The anyonic string 2d worldsheet is the 2d surface topological operator.\footnote{Of course we know that in above 3d spacetime (such as 4d and 5d that we concern),
a 0d particle by itself can only have bosonic or fermionic statistics. A 0d particle does not have fractional or anyonic statistics \cite{Wilczek1990BookFractionalstatisticsanyonsuperconductivity} in above 3d spacetime.
However, 1d worldine and 2d worldsheet can be linked in 4d.
Triple and quadruple 2d worldsheets can be linked in 4d, etc (See \cite{WangLevin1403.7437, Jiang1404.1062, Wang1404.7854} and \cite{CWangMLevin1412.1781, 
Putrov2016qdo1612.09298PWY, 1602.05951WWY, Wang2019diz1901.11537}).
These give rise to ``anyonic statistics'' to multi excitations of gapped particles or gapped strings in a 4d spacetime and above.  
}

In summary, based on the anomaly {cancellation} and cobordism constraints, 
we propose that the SM and Georgi-Glashow $su(5)$ GUT
(with 15 Weyl fermions per generation, and with a discrete 
baryon minus lepton number $\Z_{4,X}$ preserved) contains a new hidden sector 
that can be a linear combination of above Scenarios \cite{JW2006.16996,JW2008.06499}.
In particular, we can focus on the {new hidden sectors} given by a 4d TQFT (with the anomaly index $\upnu_{\rm 4d}$)
and a 5d iTQFT (with the 4d boundary's anomaly index $\upnu_{\rm 5d}$),
so \eq{eq:anomaly} becomes
\bea \label{eq:anomaly-match}
\boxed{
  (-{(N_{\text{gen}}=3)}+ 
 ( \sum_{j=e,\mu,\tau,\dots }n_{\nu_{j,R}}) + \upnu_{\rm 4d} - \upnu_{\rm 5d}) = 0 \mod 16.
}
\eea

\subsection{Symmetry Breaking vs Symmetry Extension:\\
Dirac or Majorana masses vs Topological mass}

{The distinctions between  {\bf Dirac mass, Majorana mass, and Topological mass} are already explored in  \Refe{JW2006.16996}.}
They represent the Scenarios \ref{Dirac}, \ref{Majorana}, and  \ref{Z4XTQFT} respectively in \Sec{sec:Consequences}. 
Here we summarize their essences that:\\[2mm]
\noindent 
$\bullet$ {\bf Symmetry breaking}: Dirac mass and Majorana mass are induced by \emph{symmetry breaking} --- either 
global symmetry breaking or gauge symmetry breaking, for example via the Anderson-Higgs mechanism or through Yukawa-Higgs term.
{More precisely, we start from a symmetry group (specifically here an internal symmetry, global or gauged) $G$, and we break $G$ down to
an appropriate subgroup $G_{\text{sub}} \subseteq G$ to induce quadratic mass term for matter fields. Mathematically we write 
an \emph{injective} homomorphism ${\iota}$:
$$
G_{\text{sub}} \overset{\iota}{\longrightarrow} G.
$$
For example, in Anderson-Higgs mechanism, for a Bardeen-Cooper-Schrieffer type $\Z_2$-gauged superconductor, we have
$G_{\text{sub}}=\Z_2$ and $G=\U(1)$ electromagnetic gauge group.
For the SM electroweak Higgs mechanism, we have $G_{\text{sub}}=\frac{\SU(3) \times \U(1)_Y}{\Z_{\gcd(q,3)}}$ 
and $G = G_{\SM_q} \equiv \frac{\SU(3) \times   \SU(2) \times \U(1)_{\text{EM}}}{\Z_q}$ with $q=1,2,3,6$ and the appropriate greatest common divisor (gcd).
}\\[2mm]
\noindent
$\bullet$ {\bf Symmetry extension}: Topological mass as the energy gap above a TQFT with 't Hooft anomaly
(of a spacetime-internal symmetry $G$) can be induced by \emph{symmetry extension} \cite{Wang2017locWWW1705.06728}. 
The {symmetry extension} mechanism extended the original Hilbert space (with nonperturbative global anomalies)
to an enlarged Hilbert space by adding extra degrees of freedom to the original quantum system 
{(see \cite{Prakash2018ugo1804.11236,PrakashJW2011.13921} for explicit quantum Hamiltonian lattice constructions)}.
The enlarged Hilbert space is meant to trivialize the 't Hooft-anomaly in $G$ in an extended $\tilde G$.
The pullback $r^*$ of the map
$$
\tilde G \overset{r}{\longrightarrow} G
$$
can be understood as part of the group extension in an exact sequence \cite{Wang2017locWWW1705.06728}, while 
in general this can be generalized as the fibrations of their classifying spaces and higher classifying spaces 
\cite{Tachikawa2017gyf1712.09542, Wang1801.05416, Wan2018djlW2.1812.11955}. {In many simplified cases, we
have a \emph{surjective} homomorphism ${r}$ in a short-exact sequence of group extension:
$$
1 \longrightarrow N_{\text{normal}} \longrightarrow \tilde G \overset{r}{\longrightarrow} G \longrightarrow 1,
$$
where $G= \frac{\tilde G}{ N_{\text{normal}}}$ becomes a quotient group of $\tilde G$ whose normal subgroup is $N_{\text{normal}}$.
The $G$-anomaly becoming anomaly-free in the extended ${\tilde G}$ requires the essential use of algebraic topology criteria, 
such as the Lydon-Hochschild-Serre spectral sequence method \cite{Wang2017locWWW1705.06728}.
We will explain further details in \Sec{sec:sym-ext}.
}

\subsection{Gauging a discrete Baryon {\bf B}, Lepton {\bf L}, and Electroweak Hypercharge $Y$}

We provide some more logical motivations why we should preserve $\Z_{4,X}$ and dynamically gauge $\Z_{4,X}$ with $X \equiv
5({ \mathbf{B}-  \mathbf{L}})-4Y$, at a higher energy
(Scenario \ref{Z45dSET} and \ref{Z45dSET2} in \Sec{sec:Consequences}):

\begin{enumerate}[leftmargin=-0mm]

\item First, as stated before, the $\Z_{4,X}$ is a good global symmetry read from the quantum numbers of SM particles and SM path integral kinematically.
It is a global symmetry that has \emph{not yet} been dynamically gauged in the $G_{\text{SM}_q}$ nor in the SU(5) of the $su(5)$ GUT.

\item The $\Z_{4,X} = Z(\Spin(10))$ sits at the center subgroup $\Z_{4}$ of the Spin(10) for the $so(10)$ GUT \cite{GarciaEtxebarriaMontero2018ajm1808.00009}.
Thus the $\Z_{4,X}$ must be dynamically gauged, if the $so(10)$ GUT is a correct path to unification \emph{at a higher energy}.
It is natural to consider the following group embedding from the GUT to the SM \cite{WW2019fxh1910.14668, JW2006.16996,JW2008.06499}:
\bea \label{eq:embedding}
\hspace{-10mm}
 {\frac{\Spin(d) \times
\Spin(10)}{{\Z_2^F}} 
\supset 
\Spin(d) \times_{\Z_2^F} \Z_{4,X}\times \SU(5) 
\supset 
\Spin(d) \times_{\Z_2^F} \Z_{4,X}\times \frac{\SU(3) \times   \SU(2) \times \U(1)}{\Z_6}}.
\eea
\noindent
$\bullet$ It is worthwhile mentioning that the $\Z_{16}$ global anomaly
(occurred in the cobordism group for the $G=\Spin(d) \times_{\Z_2^F} \Z_{4,X}\times \SU(5)$ and $\Spin(d) \times_{\Z_2^F} \Z_{4,X}\times G_{\SM_q}$)
{disappears} in the case of $G=\frac{\Spin(d) \times
\Spin(10)}{{\Z_2^F}}$.

So the $\Z_{4,X}$ can be an anomalous symmetry
in the ${\SM_q}$ and the $su(5)$ GUT,
but the $\Z_{4,X}$ is an anomaly-free symmetry in the $so(10)$ GUT.\footnote{The anomalous symmetry 
means a non-onsite symmetry in the condensed matter terminology, that cannot be
realized acting only locally on a 0-simplex (a point).
The anomaly-free symmetry means an onsite symmetry in the condensed matter terminology, that acts only locally on a 0-simplex (a 0d point), which can be easily gauged
by coupling \cred{to} dynamical variables living on 1-simplices (1d line segments).
}

\noindent
$\bullet$ The 15n Weyl fermion SM$_q$ or the 15n Weyl fermion $su(5)$ GUT alone may have a $\Z_{16}$ global anomaly,
while they can become anomaly-free at a higher-energy 16n Weyl fermion $so(10)$ GUT \cite{WangWen2018cai1809.11171}. 
This fact motivates \Refe{JW2008.06499} to propose an analogous concept of topological quantum phase transition 
happens between two energy scales:
(1) above the energy scale of the SM$_q$ or the $su(5)$ GUT, (2) below the energy scale of the $so(10)$ GUT.
(See the alternative interpretations of topological quantum phase transition by deforming the SM in \Sec{label:Summary}.)

\item Global symmetry must be gauged or broken in quantum gravity. If for the above reasons, we 
ask the $\Z_{4,X}$-symmetry to be preserved, then the $\Z_{4,X}$ must be dynamically gauged at a higher energy
for the sake of quantum gravity.\footnote{String theory landscape and swampland program develops the similar concepts 
of the use of cobordism for quantum gravity, see \cite{McNamara2019rupVafa1909.10355} and References therein.
This can be understood as the deformation classes of quantum gravity.
The deformation classes of quantum field theory is also proposed by Seiberg in \cite{NSeiberg-Strings-2019-talk}.
In our context, we propose that the whole quantum system including the low energy SM plus additional hidden sectors,
must correspond to the trivial group element $0$ class 
in $\TP_{d=5}(\Spin \times_{\Z_2^F} \Z_{4,X} \times G_{\SM_q})$.
Similarly, the whole quantum system including the $su(5)$ GUT plus additional hidden sectors,
must correspond to the trivial group element $0$ class 
in $\TP_{d=5}(\Spin \times_{\Z_2^F} \Z_{4,X} \times {\SU(5)})$.\\
$\bullet$ \emph{Anomaly matching}:  
To take a step back, for a usual quantum field theory, one can try to match the index $\upnu$ of 't Hooft anomaly of ultraviolet high energy (UV) with infrared low energy (IR).
The index  $\upnu$ is a renormalization group (RG) flow invariant but possibly can be nonzero. 
This is the \emph{anomaly matching} of the index $\upnu$ between UV and IR theories. \\
$\bullet$ \emph{Anomaly cancellation}: Here in contrast, in our case, we consider the whole quantum system (low energy and high energy) into account, 
due to our assumption that the $\Z_{4,X}$ is preserved thus gauged at the quantum gravity scale,
we must have the system \emph{anomaly matched} to a trivial group element with the total index $\upnu = 0$ in the cobordism class 
(similar to  \cite{McNamara2019rupVafa1909.10355}). 
We may also quote this cancellation as the anomaly matching to zero.\\[2mm]
{If we ignore the dynamical gravity,  
we can make some comments about the UV completion of Ultra Unification with a local-tensor product Hilbert space (namely, as a regularized quantum lattice model):\\
$\bullet$ If the $\Z_{4,X} \supseteq \Z_2^F$ is treated as a global internal symmetry (thus not dynamically gauged), the Ultra Unification requires a fermionic Hilbert space
with local gauge-invariant fermionic operators such that the $\Z_{4,X}  \supseteq \Z_2^F$ acts onsite in the 4d-5d coupled system. 
Namely, the system can be defined on
a manifold with a fermionic $\Spin \times_{\Z_2^F} \Z_{4,X}$-structure with its $\Z_{4,X}$ generator square $X^2=(-1)^F$ as the fermion parity.\\
$\bullet$ If the $\Z_{4,X} \supseteq \Z_2^F$ is dynamically gauged at higher energy (as it should be),
the Ultra Unification requires a bosonic Hilbert space
with only local gauge-invariant bosonic operators in the 4d-5d coupled system, without any global symmetry.
Namely, the system can be defined on
a manifold with a bosonic $\SO$-structure.
\\[2mm]
Overall, to have the Ultra Unification applied to our Universe's quantum vacuum, 
it is convenient to regard the $\Z_{4,X}$ looks like a global symmetry (thus a fermionic Hilbert space) at a lower energy around SM scales,
but the $\Z_{4,X}$ is eventually dynamically gauged at higher energy (thus eventually a bosonic Hilbert space at the deep UV completion). 
Moreover, since the $\Spin \times_{\Z_2^F} \Z_{4,X}$ gauge field
is not an ordinary abelian discrete gauge field but with the extra constraint $w_2(TM) = \CA_{{\Z_2}}^2$,
we may require to sum over the $\Spin \times_{\Z_2^F} \Z_{4,X}$-gauge bundle and $\Spin \times_{\Z_2^F} \Z_{4,X}$-gauge field $\CA_{{\Z_4}}$ altogether properly.
}} 
 
\end{enumerate}

\subsection{Topological Phase Sector and Topological Force}
\label{sec:TopologicalSectorForce}
Topological Force and the gapped Topological Phase Sector have specific physical and mathematical meanings in our context.
We should clarify what they are, and then what they are not in the next:\\[-8mm]
\begin{enumerate}[leftmargin=.mm, label=\textcolor{blue}{\arabic*}., ref={\arabic*}]

\item {\bf Topological Phase Sector}: In \Sec{sec:Consequences} and \eq{eq:anomaly-match}, 
we propose a Topological Phase Sector beyond the Standard Model (BSM)
includes an appropriate linear combination of the following theories (selecting the best scenario to fit into the HEP phenomenology):

\begin{enumerate}[leftmargin=4.mm, label=\textcolor{blue}{(\alph*)}., ref={(\alph*)}]
\item {\bf 4d long-range entangled gapped topological phase}
with an energy gap (named the gap $\Delta_{\rm TQFT}$) whose low energy physics is characterized by a {\bf 4d noninvertible 
topological quantum field theories}. This 4d TQFT is a Schwarz type unitary TQFT (which is the 4d analog of the 3d Chern-Simons-Witten theories \cite{Schwarz1978cn, Witten1988hfJonesQFT}).

$\bullet$ This 4d TQFT has a 't Hooft anomaly \cite{tHooft1979ratanomaly} of a global symmetry $G$.
We named the anomaly index $\upnu_{\rm 4d}$ for this 4d TQFT.

$\bullet$ Proper mathematical tools to study this 4d TQFT requires the category or higher category theories.

\item {\bf  5d short-range entangled gapped topological phase}
with an energy gap 
whose low energy physics is characterized by a {\bf 5d invertible 
topological quantum field theory (iTQFT)}. This 5d iTQFT is also a unitary TQFT. 
But the iTQFT is nontrivial distinct from a trivial gapped vacuum 
only in the presence of a global symmetry 
$G$, see Footnote \ref{ft:Conventions}. 
A $G$-symmetric iTQFT
is mathematically given by a $G$-cobordism invariant, classified by an appropriate cobordism group
$\Omega^{d}_{G} \equiv
\TP_d(G)$, defined in the Freed-Hopkins classification of invertible topological phases (TP) \cite{Freed2016}.

$\bullet$ The boundary of this 5d iTQFT has a 4d 't Hooft anomaly of a global symmetry $G$.
We named the anomaly index $\upnu_{\rm 5d}$ for this 5d iTQFT.

$\bullet$ Proper mathematical tools to study this 4d TQFT requires characteristic classes, cohomology, and cobordism theories.

\item {\bf  5d long-range entangled gapped topological phase}
with an energy gap
whose low energy physics is characterized by a {\bf 5d TQFT}. 
This is the case when the discrete $X$ symmetry is dynamically gauged, 
stated in Scenario \ref{Z45dSET} and \ref{Z45dSET2} in \Sec{sec:Consequences}.

\end{enumerate}

\item {\bf Topological Force}:
In the context of \Sec{sec:Consequences}, 
Topological Force is a discrete gauge force mediated between the linked worldvolume trajectories 
(1d worldlines, 2d worldsheets from gapped extended operators) 
via fractional or categorical statistical interactions (See Sec.~5 and 6 of \cite{JW2006.16996}).\\

$\bullet$ {\bf  Bosonic finite group gauge theory}: 
The conventional discrete gauge theories are {\bf  bosonic} types of finite group gauge theories \cite{KraussWilczekPRLDiscrete1989,Dijkgraaf1989pzCMP}.
Bosonic types mean that their ultraviolet (UV) completion only requires a local tensor product Hilbert space of local (gauge-invariant) bosonic operators;
the UV completion does not require local (gauge-invariant) fermionic operators.
The underlying TQFT does not require the spin structures and can be defined on non-Spin manifolds (such as the oriented SO structures). 
The TQFTs are known as bosonic or non-Spin TQFTs.\\

$\bullet$ {\bf  Fermionic finite group gauge theory}: The discrete gauge theories for our gapped Topological Phase Sectors for the beyond SM hidden sector
are {\bf  fermionic} types of finite group gauge theories \cite{Wang1801.05416, GuoJW1812.11959}.
Fermionic types mean that their UV completion must require local (gauge-invariant) fermionic operators.
The underlying TQFT requires the additional spin structures defined on Spin manifolds. The TQFTs are known as Spin TQFTs.
In fact the 4d TQFT in \Sec{sec:Consequences} requires the $\Spin  \times_{\Z_2^F} \Z_{4,X}$ structure and can be defined on
the $\Spin \times_{\Z_2^F} \Z_{4,X}$ manifolds (including both Spin manifolds and some non-Spin manifolds). 

\end{enumerate}

We should emphasize that our {Topological Phase Sector and Topological Force}
are \emph{not} the kinds of 
Chern class topological terms which are already summed over in the continuous Lie group gauge theory.
Namely, our {Topological Phase Sector and Topological Force} are {\bf not} the followings:

\noindent
--- {\bf The $\theta$-term with or without a dynamical $\theta$-axion} \cite{Weinberg1977axion,Wilczek1977axion},
well-known as $\theta F \wedge F$ or $\theta F\tilde{F}$ in the particle physics,
is in fact related to the  {\bf second Chern class} $c_2(V_{G})$ and the square of the  {\bf first Chern class} $c_1(V_{G})$
of the associated vector bundle of the gauge group $G$:
\bea
  \frac{\theta}{8\pi^2} \Tr( {F}\wedge  {F}) =\frac{\theta}{2} c_1(V_{G})^2 - \theta\; c_2(V_{G}).
\eea 
In particular, here we consider $G$ as the U(N) or SU(N) gauge group,
so we can define the Chern characteristic classes associated with complex vector bundles.
The $V_{G}$ is the associated vector bundle of the principal $G$ bundle.
This  $\theta$-term is a topological term, but it is summed over as a weighted factor to define a 
Yang-Mills gauge theory partition function \cite{AharonyASY2013hdaSeiberg1305.0318, Gaiotto2017yupZoharTTT,Wan2019oyr1904.00994}. 
This  $\theta$-term does {\bf not} define a quantum system or a quantum phase of matter by itself,
distinct from our 4d TQFT (with intrinsic topological order) and 5d iTQFT (with SPTs) as certain unitary quantum phases of matter by themselves.

\noindent
---  {\bf  The instantons} \cite{BelavinBPST1975, tHooft1976instanton} or {\bf  the sphalerons} \cite{KlinkhamerMantonsphalerons1984},
are also not the {Topological Phase Sector and Topological Force} in our context.
Instantons and sphalerons are again the objects with nontrivial Chern class integrated over the spacetime manifold.
Those objects are already defined as part of the SM and GUT continuous group gauge theories.

As we will mention in \Sec{sec:SMGUTPathIntegral}, we can also include
(1) the $\theta$-term with or without a dynamical $\theta$-axion,
(2) instantons, and (3) sphalerons, into the 
{Standard Model and the $su(5)$ GUT path integral.}
These objects are already in the old paradigm of the SM and GUT models.
These objects do not affect the anomaly {cancellation} 
and cobordism constraints (especially the $\Z_{16}$ global anomaly) 
discussed in \Sec{sec:Anomaly}.\footnote{However, those $\theta$ terms in Yang-Mills gauge theory may affect the higher anomalies 
involving higher generalized global symmetries, 
see for example \cite{Gaiotto2017yupZoharTTT,Wan2019oyr1904.00994}.} 
These objects belong to the {Standard Model and the $su(5)$ GUT path integral (\Sec{sec:SMGUTPathIntegral})},
not to the Topological Phase Sector (TQFT) path integral (\Sec{sec:TQFTPathIntegral}),
but they all can be included as part of the Ultra Unification path integral (\Sec{sec:PathIntegral}).

\section{Ultra Unification Path Integral}
\label{sec:PathIntegral}

In this section, we provide the functional path integral (i.e., partition function) 
$\bZ_{\UU}$ of Ultra Unification, 
which includes the standard paradigm of the Standard Model path integral $\bZ_{\SM}$
or the Georgi-Glashow $su(5)$ GUT path integral $\bZ_{\GUT}$
in \Sec{sec:SMGUTPathIntegral}. Then we provide the Topological Phase Sector TQFT
path integral $\bZ_{\TQFT}\equiv {\bf Z}_{\text{5d-iTQFT}} \cdot {\bf Z}_{\text{4d-TQFT}}$
in \Sec{sec:TQFTPathIntegral}

\subsection{Standard Model and the $su(5)$ GUT 
Path Integral coupled to $X \equiv 5({ \mathbf{B}-  \mathbf{L}})-4Y$}
\label{sec:SMGUTPathIntegral}

\subsubsection{Standard Model Path Integral coupled to $X$}
Now we describe the Standard Model (SM) path integral {in the Minkowski (or Lorentz) signature}:  
\bea \label{eq:ZSM}
\bZ_{\SM}[\CA_{{\Z_4}}]\equiv
\int [{\cal D} {\psi}] [{\cal D}\bar{\psi}] [{\cal D} A][{\cal D} \phi] \dots
\exp( \ii \left. S_{\text{SM}}[\psi, \bar{\psi}, A, \phi, \dots,  \CA_{\Z_4}] \right  \rvert_{M^4}).
\eea
The $\dots$ depends on the details of which variant versions of SM that we look at (e.g., adding axions or not).
In the schematic way, we have the action  {$S$}:
\begin{multline} \label{eq:SSM}
S_{\text{SM}}= \int_{M^4} 
{\Big(\Tr(F_{I} \wedge \star F_{I})
- \frac{\theta_I}{8 \pi^2} {g}_I^2
 \Tr( F_{I} \wedge F_{I} ) \Big)}
+ \int_{M^4} \Big(\bar{\psi} (\ii \slashed{D}_{A,\cA_{\Z_4}}) \psi\\
+ | {D}_{\mu, A, \cA_{\Z_4}}\phi |^2 -{\rm U}(\phi)
 -( {\psi}^\dagger_L \phi \psi_R +{\rm h.c.}) \Big)\,\dd^4 x.
\end{multline}
But more precisely we really need more details in the Lagrangian  {$\cL$} with Weyl fermions, with $S \equiv \int \cL \dd^4x$:  
\begin{multline} \label{eq:LSM}
\cL_{\rm SM}= 
\cL_{\text{YM}}
+\cL_{\theta\text{-Chern}} 
+\cL_{\text{Weyl}}
+\cL_{\text{Higgs}}
+\cL_{\text{Yukawa-Higgs}}
\\= \cred{\sum_{I=1,2,3}}
 -\frac{1}{4} F_{I,\mu\nu}^\ra F_{I}^{\ra \mu\nu}
 {-\, \frac{{\theta_I}}{{64} \pi^2} {g}_I^2 \epsilon^{\mu\nu \mu'\nu'} F_{I,\mu\nu}^\ra F_{I, \mu'\nu'}^\ra}
+ {\psi}^\dagger_L  (\ii \bar  \sigma^\mu {D}_{\mu, A, \cA_{\Z_4}} ) \psi_L
 + {\psi}^\dagger_R  (\ii  \sigma^\mu {D}_{\mu, A, \cA_{\Z_4}} ) \psi_R
 \\
+ | {D}_{\mu, A, \cA_{\Z_4}}\phi |^2 -{\rm U}(\phi)
 -( {\psi}^\dagger_L \phi \psi_R +{\rm h.c.}).
\end{multline}
Here come some Remarks:
\begin{enumerate}[leftmargin=.mm, label=\textcolor{blue}{\arabic*}., ref={\arabic*}]
\item \label{lYM}
Yang-Mills gauge theory \cite{PhysRev96191YM1954} has the action
$S_{\text{YM}}=\int \Tr(F_{} \wedge \star F_{})$ and Lagrangian $\cL_{\text{YM}}= -\frac{1}{4} F_{\mu\nu}^\ra F^{\ra{\mu\nu}} $.\\
The $F$ is the Lie algebra valued field strength curvature 2-form
$F = \dd A - \ii \cred{ g} A \wedge A$, with its Hodge dual \cred{$\star F$}, all written in differential forms.
In the trace ``Tr'' we pick up a Lie algebra representation ${\bf R}$ whose \cred{Lie algebra} generators \cred{$\rT^a$} labeled by ``$\ra$.''
{We have also the subindex $I=1,2,3$ to specify the SM Lie algebra sectors $u(1)$, $su(2)$, or $su(3)$.}

{{More precisely 
$F=\frac{1}{2} F_{\mu\nu}( \dd x^\mu \wedge \dd x^\nu) =\frac{1}{2} F^a_{\mu\nu} \rT^a( \dd x^\mu \wedge \dd x^\nu)$,
and we
define the commutator $[\rT^b, \rT^c]=\ii f^{bcd} \rT^d$ with a structure constant $f^{bcd}$, 
then $F^a_{\mu\nu} = \partial_\mu A^a_\nu -\partial_\nu A^a_\mu +  g f^{bca} A^b_{\mu} A^c_{\nu}$.
{Note} that $\Tr(\rT^a \rT^b)=C({\bf R}) \delta^{ab}$ for some constant of representation ${\bf R}$.
Here for the fundamental representation ${\bf R}$, we take $\Tr(\rT^a \rT^a)=\frac{1}{2}$.
Then we have $\Tr(F \wedge * F)
=(-1)^{\rm{s}}
\frac{1}{2}
 \Tr(F_{\mu\nu} F^{\mu\nu} ) \dd^4x
=(-1)^{\rm{s}}
(\frac{1}{4} )
F^a_{\mu\nu} (F^a)^{\mu\nu} \dd^4x$ 
with the $(-1)^{\rm{s}}$ as the sign of the determinant of the spacetime metric.
Here $(-1)^{\rm{s}}=-1$ in the Minkowski signature.
Notice that here we normalize the $u(1)$ Yang-Mills in \eq{eq:SSM} and \eq{eq:LSM} slightly differently from the conventional $u(1)$ Maxwell theory
by scaling a factor $\Tr(1)$ of the $u(1)$ by a $\frac{1}{2}$.
}}

The $\theta$-term and dynamical $\theta$-axion: We can also introduce the Chern class topological $\theta$-term
\cred{$S_{\theta\text{-Chern}}=-\int\frac{\theta}{8 \pi^2}g^2
 \Tr( F \wedge F )$} and  \cred{$\cL_{\theta\text{-Chern}}=-\frac{\theta}{{64} \pi^2} g^2 \epsilon^{\mu\nu \mu'\nu'} F_{\mu\nu}^\ra F_{\mu'\nu'}^\ra$}.
Given a ${\U(\rN)}$ or SU(N) bundle $V_{G}$ and its field strength $\widehat{F}$,
the first and second Chern classes are given by
$c_1(V_{G})= \frac{\Tr {F}}{2\pi}$
and $c_2(V_{G})= -\frac{1}{8\pi^2} \Tr({F}\wedge {F}) + \frac{1}{8\pi^2} (\Tr {F}) \wedge(\Tr {F})$, so that
$\frac{\theta}{8\pi^2} \Tr( {F}\wedge  {F}) =\frac{\theta}{2} c_1(V_{G})^2 - \theta\; c_2(V_{G})$.
If a dynamical $\theta$-axion \cite{Weinberg1977axion,Wilczek1977axion} is introduced,
it requires a summation of the compact $\theta$ in the path integral measure
$\int [{\cal D} \theta]$.

\cred{There is an overall constant that can be absorbed into the field $A$ and coupling $g$ redefinition.\footnote{\cred{For example,
by redefining $A \to A'= \frac{1}{g}A$ and $F \to F'= \frac{1}{g}F$, then 
$F'= \dd A' - \ii  A' \wedge A'$
and 
$F'^a_{\mu\nu} = \partial_\mu A'^a_\nu -\partial_\nu A'^a_\mu +  f^{bca} A'^b_{\mu} A'^c_{\nu}$.
Then we can also write 
$$S_{\text{YM}}+S_{\theta\text{-Chern}}=
\frac{1}{g^2}\int \Tr(F'_{} \wedge \star F'_{})
-\int\frac{\theta}{8 \pi^2}
 \Tr( F' \wedge F'), \text{ and } \cL_{\text{YM}}+\cL_{\theta\text{-Chern}}= -\frac{1}{4 {g^2}} {F'}_{\mu\nu}^\ra {F'}^{\ra{\mu\nu}} 
-\frac{\theta}{{64} \pi^2}  \epsilon^{\mu\nu \mu'\nu'} {F'}_{\mu\nu}^\ra {F'}_{\mu'\nu'}^\ra.$$
}
}}

The path integral $\int [{\cal D} A]$ for continuous Lie group gauge field theory (here U(1), SU(N), U(N) for the SM and $su(5)$ GUT),
really means (1) the summation of all inequivalent principal gauge bundles $P_A$, and 
then (2) the summation of all inequivalent gauge connections $\tilde A$ (under a given specific principal gauge bundles $P_A$),
where $\tilde A$ is a (hopefully globally defined physically) $1$-form gauge connection.
So we physically define:
$$
\int [{\cal D} A] \dots \equiv \sum_{\text{gauge bundle $P_A$}}
\int [{\cal D} \tilde A] \dots.
$$

\item \label{lDirac}
Dirac fermion theory has
$S_{\text{Dirac}}= \int \bar{\psi} (\ii \slashed{D}_{A}) \psi \,\dd^4 x$ and
$\cL_{\text{Dirac}}= \bar{\psi} (\ii \slashed{D}) \psi$ with 
the Dirac spinor $\psi$ defined as
a section of the spinor bundles.
The $\bar{\psi} (\ii \slashed{D}) \psi$ is an inner product in the complex vector space
with the Dirac operator $\slashed{D}_{A}$ as a natural linear operator in the vector space.
The path integral 
$\int [{\cal D} {\psi}] [{\cal D}\bar{\psi}]$
is (1) the summation of all inequivalent spinor bundles, and 
then (2) the summation of all inequivalent sections (as spinors) of spinor bundles (under a given specific
spinor bundle). We requires the spin geometry and spin manifold, in particular 
we require the $\Spin \times_{\Z_2} \Z_4 =\Spin \times_{\Z_2^F} \Z_{4,X}$ structure. 

In fact preferably we present not in the Dirac spinor basis, but we present all of \eq{eq:LSM} in the Weyl spinor basis (below).

\item \label{lWeyl}
Weyl fermion theory has
$S_{\text{Weyl}}=\int \cL_{\text{Weyl}} \,\dd^4 x= \int {\psi}^\dagger_L  (\ii \bar  \sigma^\mu {D}_{\mu, A, \cA_{\Z_4}} ) \psi_L\,\dd^4 x$. 
Weyl spinor bundle splits the representation of the Dirac spinor bundle.
Weyl spinor again is defined as the section of Weyl spinor bundle.
The Weyl spacetime spinor is in
${\bf 2}_L$ {of} ${{\Spin(1,3)}}=\rm{SL}(2,\C)$  with a complex representation in the Lorentz signature,
or ${\bf 2}_L$  {of} $\Spin(4)=\SU(2)_L \times \SU(2)_R$ with a {pseudoreal representation}
in the Euclidean signature. 
We also write the analogous right-handed Weyl fermion theory.
The $\sigma^\mu$ and $\bar{\sigma}^\mu$ are the standard spacetime spinor rotational $su(2)$ Lie algebra generators.
We will emphasize and illuminate the meanings of covariant derivative ${D}_{\mu, A, \cA_{\Z_4}}$ altogether in Remark \ref{lgauge}.

\item \label{lHiggs}
Higgs theory has
$S_{\text{Higgs}} = \int \cL_{\text{Higgs}} \,\dd^4 x= \int   \Big( | {D}_{\cred{\mu, A, \cA_{\Z_4}}}\phi |^2 -{\rm U}(\phi)\Big)\,\dd^4 x$ 
 with gauged kinetic and potential terms.
%
The Higgs field bundle is typically a trivial complex line bundle.\footnote{In Higgs theory, people in general do not consider
nontrivial complex line bundles for Higgs field. But it may be amusing to consider the alternative.}
The Higgs scalar field is the section of a field bundle.
The electroweak Higgs is in complex value $\C$ and also in ${\bf 2}$ of SU(2) gauge field.
Again by doing summation 
$\int [{\cal D} \phi]$
we (1) sum over the field bundles,
and (2) sum over the section of each field bundle.
We illuminate the meanings of covariant derivative ${D}_{\mu, A, \cA_{\Z_4}}$ altogether in Remark \ref{lgauge}.
 
\item \label{lgauge}
Covariant derivative operator ${D}_{\mu, A, \cA_{\Z_4} }$ in \eq{eq:LSM} is defined as:
\bea
{D}_{\mu, A, \cA_{\Z_4} } \equiv \nabla_{\mu} - \ii g \,  q_{{\bf R}} \,  A_\mu - \ii q_{{X}} \cA_{\Z_4, \mu}.
\eea
Placed on a curved spacetime (with a non-dynamical metric, only with background gravity)
requires a covariant derivative $\nabla_{\mu}$, and a spin connection for the spinors.\\
Comments about the term $g \,  q_{{\bf R}} \,  A$ in a differential form (e.g., quantum numbers read from Table 1 in \cite{JW2006.16996}):\\
\bea
g  \,  q_{{\bf R}}\,  A \equiv (q_e  A_{{u(1)},\mu} 
+ g_{su(2)}  \sum_{\ra=1}^3\frac{\varsigma^\ra}{2}   A_{{su(2)},\mu}^\ra
+ g_{su(3)} \sum_{\ra=1}^8\frac{\tau^\ra}{2} A_{{su(3)},\mu}^\ra )\dd x^\mu.
\eea
The ${\varsigma^\ra}$ and ${\tau^\ra}$ are the rank-2 and rank-3 Lie algebra generator matrix representations for 
${su(2)}$ and ${su(3)}$ respectively.  
The ${D}_{\mu, A, \cA_{\Z_4}}$ acting on $\psi_L$ contains the $su(2)$ gauge field.
The ${D}_{\mu, A, \cA_{\Z_4}}$ acting on $\psi_R$ does not contain the $su(2)$ gauge field, 
because the $su(2)$ weak interaction is a maximally parity violating chiral gauge theory.
The ${D}_{\mu, A, \cA_{\Z_4}}$ acting only on quarks (both $\psi_L$ and $\psi_R$) contain the $su(3)$ gauge field. 
The ${D}_{\mu, A, \cA_{\Z_4}}$ acting on $\phi$ contains the $su(2) \times u(1)$ gauge field. 
 \\

Comments about the term $q_{{X}} \cA_{\Z_4}$ (e.g., quantum numbers read from Table 1 and 2 in \cite{JW2006.16996}):\\
$\bullet$ The ${D}_{\mu, A, \cA_{\Z_4}}$ acts on all left-handed SM Weyl fermion $\psi_L$ via its $\cA_{\Z_4}$ charge $q_{{X}}=1$. \\
$\bullet$ The ${D}_{\mu, A, \cA_{\Z_4}}$ acts on all right-handed SM Weyl fermion $\psi_R$ via its $\cA_{\Z_4}$ charge $q_{{X}}=-1$.\\ 
$\bullet$ The ${D}_{\mu, A, \cA_{\Z_4}}$ acts on the electroweak Higgs $\phi$ via its $\cA_{\Z_4}$ charge $q_{{X}}=2$. 

The subtle part is that $\cA_{\Z_4}$ should be treated as a cohomology class, such as a cohomology or cochain gauge field.
The $\CA_{{\Z_4}} \in \H^1(M,\Z_4)$ is the generator from  $\H^1(\B\Z_{4,X},\Z_4)$.
In physics, for the continuum QFT theorists who prefer to think $\Z_{4,{X}}  \subset \U(1)_{X}$
as a continuum gauge field breaking down to a discrete $\Z_{4,{X}}$,
we can introduce an extra $\Z_4$ charge new Higgs field $\varphi$ and its potential ${\rm V}(\varphi)$:
\bea
{\psi}^\dagger_L  (\ii \bar  \sigma^\mu {D}_{\mu,  \cA_{\Z_4}} ) \psi_L
 + {\psi}^\dagger_R  (\ii  \sigma^\mu {D}_{\mu, \cA_{\Z_4}} ) \psi_R
 + |(\partial_\mu - \ii 2 \cA_{\Z_4, \mu} )\phi|^2 +|(\partial_\mu - \ii 4 \cA_{\Z_4, \mu})\varphi|^2 + {\rm V}(\varphi) +\dots.
\eea
There are extra superconductivity-like term $\phi^2 \varphi^\dagger+ (\phi^\dagger)^2 \varphi$ does not break the $\Z_4$.
Their $\Z_4$ or U(1) transformations are:
$$
\psi_L \to \psi_L \e^{\ii  \frac{2\pi}{4}},\;\;
\psi_R \to \psi_R \e^{-\ii  \frac{2\pi}{4}},\;\;
\phi \to \phi \e^{\ii 2 \frac{2\pi}{4}},\;\;
\varphi \to \varphi.$$
$$
\psi_L \to \psi_L \e^{\ii  \Theta},\;\;
\psi_R \to \psi_R \e^{-\ii  \Theta},\;\;
\phi \to \phi \e^{\ii 2 \Theta},\;\;
\varphi \to \varphi \e^{\ii 4 \Theta}.$$
In the $\langle \varphi \rangle \neq 0$ condensed Higgs phase as 
a discrete $\Z_{4}$ gauge theory,
which we can dualize the theory as a level-4 BF theory \cite{Banks2010zn1011.5120}.
The formulation starts from adding $\int [D\varphi]$ in the path integral measure,
and it ends with a 2-form $\cB$ and 1-form gauge field $\cA_{\Z_4}$
\bea
\int [D\cB][D \cA_{\Z_4}] \exp(\ii \frac{ 4}{2 \pi} \int_{M^4} \cB \wedge \dd  \cA_{\Z_4}+\dots).
\eea
But more precisely, we really should formulate in terms of a cohomology/cochain TQFT and taking care of the
$\Spin \times_{\Z_2^F} \Z_{4,X}$ structure, 
which we will do in \Sec{sec:TQFTPathIntegral} (also in Sec.~5 of \cite{JW2006.16996}).

\item \label{lDirac}
{Yukawa-Higgs-Dirac} term has
$S_{\text{Yukawa-Higgs-Dirac}}= 
\int  \cL_{\text{Yukawa-Higgs-Dirac}}\,\dd^4 x
= \int   ( {\psi}^\dagger_L \phi \psi_R +{\rm h.c.})\,\dd^4 x$.
In this case, we pair the ${\psi}^\dagger_L$'s $\bar{\bf 2}$ of SU(2) with
the $\phi$'s ${\bf 2}$ of SU(2), and vice versa pair ${\psi}_L$ with $\phi^\dagger$ to get an SU(2) singlet. 
The right-handed ${\psi}_R$ here is (meant to be) an SU(2) singlet. 
This {Yukawa-Higgs-Dirac} term at the kinetic level also preserves the $\Z_{4,X}$, 
although the Higgs vacuum expectation value (vev) breaks the $\Z_{4,X}$ dynamically.

\item \label{lMajorana}
{Yukawa-Higgs-Majorana term} with Weyl fermion:\\
We can add {Yukawa-Higgs-Majorana term} for Weyl fermions. 
For example, for the left-handed $\psi_L$,
we can add a dimension-5 operator:\\ 
$S_{\text{Yukawa-Higgs-Majorana}} = \int -( {\psi}^\dagger_L \phi (\phi \ii \sigma^2 \psi_L^*) +{\rm h.c.})\,\dd^4 x
= \int -( {\psi}^\dagger_L \phi (\phi \ii \sigma^2 \psi_L^*) +{(- \psi_L^{\rm T}}\ii\sigma^2 \phi^*)  \phi^\dagger \psi_L)\,\dd^4 x$.\\
Again the $\sigma^2$ is from the $\sigma^\mu$ of the spacetime spinor rotational $su(2)$ Lie algebra generators.
Renormalizability is not an issue because we are concerned the effective field theory.
For the right-handed $\psi_R$,
we can add a dimension-3 operator for some Majorana mass coupling $M$:\\ 
$S_{\text{Yukawa-Higgs-Majorana}} = \int -M ( {\psi_R}^{\rm T} ( \ii \sigma^2)  \psi_R +{\rm h.c.})\,\dd^4 x
= \int -M( {\psi_R}^{\rm T} ( \ii \sigma^2)  \psi_R +\psi_R^\dagger (-\ii \sigma^2){\psi_R^*})\,\dd^4 x$,\\
which breaks the lepton number conservation.
However, in either cases, both {Yukawa-Higgs-Majorana terms} above break the $\Z_{4,X}$ explicitly. 
So they are not encouragingly favored if we pursue the $\Z_{4,X}$-preserving theory at least at higher energy.
\end{enumerate}

We have presented above
the {Standard Model coupled to a discrete $X$ gauge field
in the path integral \eq{eq:ZSM}, the action \eq{eq:SSM}, and the Lagrangian \eq{eq:LSM}.
Below we can quickly modify a few terms to obtain the 
$su(5)$ Grand Unification coupled to a discrete $X$ gauge field.

\subsubsection{The $su(5)$ Grand Unification Path Integral coupled to $X$}
We have the $su(5)$ GUT path integral coupled to $X$:
\bea \label{eq:Zsu5}
\bZ_{\GUT}[\CA_{{\Z_4}}]\equiv
\int {[{\cal D} {\psi}] [{\cal D}\bar{\psi}]} [{\cal D} A][{\cal D} \phi] \dots
\exp( \ii \left. S_{\text{GUT}}[\psi, \bar{\psi}, A, \phi, \dots,  \CA_{\Z_4}] \right  \rvert_{M^4}).
\eea
We should write all $\bar{\bf 5}$ 
and ${\bf 10}$ of the SU(5) as the left-handed Weyl fermions $\psi_L$, so there are
15 Weyl fermions $\psi_L$ per generation.
In \Sec{sec:Consequences}, we may or may not introduce the right-handed neutrinos here denoted as $\chi_R$.
In the schematic way, we have the action:
\begin{multline} \label{eq:Ssu5}
S_{\text{GUT}}= \int_{M^4} \Big(\Tr(F_{} \wedge \star F_{})
-\frac{\theta}{8 \pi^2} g^2
 \Tr( F_{} \wedge F_{}) \Big) 
+ \int_{M^4} \Big( {\psi}^\dagger_L  (\ii \bar  \sigma^\mu {D}_{\mu, A, \cA_{\Z_4}} ) \psi_L
 + {\chi}^\dagger_R  ({\ii}  \sigma^\mu {D}_{\mu, A, \cA_{\Z_4}} ) \chi_R
 \\
+ | {D}_{\mu, A, \cA_{\Z_4}}\phi |^2 -{\rm U}(\phi)
 -( {\psi}^\dagger_L \phi ({\ii} \sigma^2 {\psi_L'}^*) +{\rm h.c.} )
 +  \dots
  \Big)\,\dd^4 x.
\end{multline}
We are left now only with an SU(5) gauge field whose 1-form connection written as: 
\bea
g  \,  q_{{\bf R}}\,  A = 
(g_{su(5)}  \sum_{\ra=1}^{24} \cred{{\rm T}^\ra}   A_{{su(5)},\mu}^\ra
)\dd x^\mu.
\eea
We require the ${{\rm T}^\ra}$ as the rank-5 and rank-10 Lie algebra generator matrix representations for 
${su(5)}$ to couple to $\bar{\bf 5}$ and ${\bf 10}$ of SU(5) respectively.  
Yukawa-Higgs pairs the appropriate ${\psi_L}$ and ${\psi_L'}$ Weyl fermions.
We may or may not introduce the Majorana mass terms to $\chi_R$ in the $\dots$, while the consequences are already discussed 
(which break the $\Z_{4,X}$ explicitly) in the Remark \ref{lMajorana}. 
The discussions about this path integral \eq{eq:Zsu5} directly follow the above Remarks \ref{lYM}-\ref{lMajorana},
so we should not repeat.  

\subsection{Topological Phase Sector and TQFT Path Integral coupled to $X  \equiv 5({ \mathbf{B}-  \mathbf{L}})-4Y$}
\label{sec:TQFTPathIntegral}

The 15n Weyl fermion SM and $su(5)$ GUT path integrals, \eq{eq:ZSM} and \eq{eq:Zsu5}, are \emph{not} gauge invariant under the  
$\Z_{4,X}$ gauge transformation {only when the gravitational background is turned on to probe the $\Spin \times_{\Z_2^F} \Z_{4,X}$-structure via the 
$\Spin \times_{\Z_2^F} \Z_{4,X}$ gauge field $\CA_{{\Z_4}}$}.\\
$\bullet$ If the $\Z_{4,X}$ is only {coupled to} a background gauge field, this only means {the system has 't Hooft anomaly
under the $\Z_{4,X}$ anomalous symmetry} and the spacetime (Spin group) {coordinate reparametrization} 
transformations {({e.g., the Euclidean rotation or Lorentz boost part of} diffeomorphism)}.\\
$\bullet$ If the $\Z_{4,X}$ is dynamically gauged and preserved at high energy, then we must append a new sector to make the whole theory well-defined.

In any case, follow one of the scenario in \Sec{sec:Consequences}, we now provide a path integral including Topological Quantum Field Theories (TQFTs)
 to make the whole theory free from the $\Z_{16}$ global anomaly
\eq{eq:anomaly-match}:
$$
{
  (-{(N_{\text{gen}}=3)}+ n_{\nu_{e,R}} + n_{\nu_{\mu,R}} + n_{\nu_{\tau,R}} + \upnu_{\rm 4d} - \upnu_{\rm 5d}) = 0 \mod 16.
}$$

\begin{enumerate}[leftmargin=.mm, label=\textcolor{blue}{\arabic*}., ref={\arabic*}]
\item
The 5d iTQFT partition function is given by \eq{eq:5dSPT}:
\bea \label{eq:Z5diTQFT}
{{\bf Z}_{{\text{5d-iTQFT}}}^{(\upnu_{\rm 5d})}}[\CA_{{\Z_4}}]
\equiv
\exp(\frac{2\pi \ii}{16} \cdot\upnu_{\rm 5d} \cdot \eta(\text{PD}(\CA_{{\Z_2}})) \bigg\rvert_{M^5}), & \text{ with } 
{\upnu_{\rm 5d}}\in \Z_{16}, 
\quad {\CA_{{\Z_2}} \in {\rm H}^1(M,\frac{\Z_{4,X}}{\Z_2^F})}.
\eea
\item We propose the full gauge-invariant path integral, 
invariant under the mixed gauge-gravity transformation {(i.e., gauge-diffeomorphism)}
of
$\Spin \times_{\Z_2^F} \Z_{4,X}$ structure and free from its $\Z_{16}$ global anomaly as follows.\\
The SM version employs \eq{eq:ZSM} into:
\bea \label{eq:ZSMtotal}
\boxed{\bZ_{\UU} [\CA_{{\Z_4}}]
\equiv{\bf Z}_{\substack{\text{5d-iTQFT/}\\ \text{4d-SM+TQFT}}}[\CA_{{\Z_4}}]
\equiv
{\bf Z}_{{\text{5d-iTQFT}}}^{(-\upnu_{\rm 5d})}[\CA_{{\Z_4}}]
\cdot 
{\bf Z}_{{\text{4d-TQFT}}}^{(\upnu_{\rm 4d})}[\CA_{{\Z_4}}]\cdot 
\bZ_{\SM}^{(n_{\nu_{e,R}}, n_{\nu_{\mu,R}}, n_{\nu_{\tau,R}})}
[\CA_{{\Z_4}}]
.}
\eea
The GUT version employs \eq{eq:Zsu5} into:
\bea \label{eq:Zsu5total}
\boxed{\bZ_{\UU}[\CA_{{\Z_4}}]
\equiv{\bf Z}_{\substack{\text{5d-iTQFT/}\\ \text{4d-GUT+TQFT}}}[\CA_{{\Z_4}}]
\equiv
{\bf Z}_{{\text{5d-iTQFT}}}^{(-\upnu_{\rm 5d})}[\CA_{{\Z_4}}]
\cdot 
{\bf Z}_{{\text{4d-TQFT}}}^{(\upnu_{\rm 4d})}[\CA_{{\Z_4}}]
\cdot 
\bZ_{\GUT}^{(n_{\nu_{e,R}}, n_{\nu_{\mu,R}}, n_{\nu_{\tau,R}})}
[\CA_{{\Z_4}}].
}
\eea

\end{enumerate}

\subsubsection{Symmetry extension $[\Z_2] \to \Spin \times {\Z_{4,X}} \to \Spin \times_{\Z_2^F} {\Z_{4,X}}$ and a 4d 
{fermionic} {discrete} gauge theory}
\label{sec:sym-ext}

Below we ask whether we can construct a fully {gauge-diffeomorphism} invariant 5d-4d coupled partition function preserving the
${\Spin \times_{\Z_2} \Z_4}$ structure:
\bea \label{eq:5dSPT-4dTQFT-0}
{\bf Z}_{{\text{5d-iTQFT}}}^{}[\CA_{{\Z_4}}]
\cdot 
{\bf Z}_{{\text{4d-{TQFT}}}}^{}[\CA_{{\Z_4}}].
\eea
Preserving the
${\Spin \times_{\Z_2} \Z_4}$ structure means 
that under the spacetime {coordinate} background transformation {({i.e.,} diffeomorphism)} and the $\CA_{{\Z_4}}$ background gauge transformation,
the 5d-4d coupled partition function is still fully {gauge-diffeomorphism} invariant.

First, we can rewrite the 5d iTQFT partition function \eq{eq:Z5diTQFT} on a 5d manifold $M^5$ into
\begin{multline} \label{eq:5dSPT-4dTQFT}
{\bf Z}_{{\text{5d-iTQFT}}}^{(\upnu_{})}[\CA_{{\Z_4}}]
=
\exp(\frac{2\pi \ii}{16} \cdot\upnu \cdot \eta(\text{PD}(\CA_{{\Z_2}})) \bigg\rvert_{M^5})\\
= 
\exp(\frac{2\pi \ii}{16} \cdot\upnu \cdot  \Big( 8 \cdot\frac{p_1(TM)}{48}(\text{PD}(\CA_{{\Z_2}})) + 4 \cdot\text{Arf} (\text{PD}((\CA_{{\Z_2}})^3) ) 
+ 2 \cdot{\tilde{\eta}} (\text{PD}((\CA_{{\Z_2}})^4) )   + (\CA_{{\Z_2}})^5 \Big) \bigg\rvert_{M^5})\\
= 
\exp(\frac{2\pi \ii}{16} \cdot\upnu \cdot  \Big( 8 \cdot\frac{\sigma}{16}(\text{PD}(\CA_{{\Z_2}})) + 4 \cdot\text{Arf} (\text{PD}((\CA_{{\Z_2}})^3) ) 
+ 2 \cdot{\tilde{\eta}} (\text{PD}((\CA_{{\Z_2}})^4) )   + (\CA_{{\Z_2}})^5 \Big) \bigg\rvert_{M^5}),
\end{multline}
for a generic ${\upnu=-N_{\text{generation}}} \in \Z_{16}$.\\
$\bullet$ The $p_1(TM)$ is the first Pontryagin class of spacetime tangent bundle $TM$ of the manifold $M$.
Via the Hirzebruch signature theorem, we have 
$\frac{1}{3}\int_{\Sigma^4} p_1(TM) =\sigma(\Sigma^4) =\sigma$ ${=\frac{1}{8 \pi^2}\int \Tr(R(\omega) \wedge R(\omega))}$
on a 4-manifold $\Sigma^4$, where $\sigma$ is the signature of $\Sigma^4$ {while $\omega$ is the 1-connection of tangent bundle and
$R(\omega)$ is the Riemann curvature 2-form of $\omega$}.\\
So in \eq{eq:5dSPT-4dTQFT}, we evaluate the $\frac{1}{3}\int_{\Sigma^4} p_1(TM) =\sigma$ on the Poincar\'e dual (PD) of $\Sigma^4$ manifold of
the $(\CA_{{\Z_2}})$ cohomology class within the $M^5$.\footnote{{Here is a caveat:
We know that $\frac{p_1(TM)}{48}=\frac{\sigma}{16} \in \Z$ for 4d spin manifolds which makes \eq{eq:5dSPT-4dTQFT} computable. 
But we leave the precise analogous revised expression of $\frac{p_1(TM)}{48}=\frac{\sigma}{16}$ on unoriented manifolds
{(such as $\Pin^+$ manifolds) 
as a torsion class
\cite{Witten1508.04715, Witten2016cio1605.02391, 1711.11587GPW} in the parallel work \cite{PTWtoappear}.}
}
}
\\
{$\bullet$ The $\tilde{\eta}$ is a mod 2 index of 1d Dirac operator as a cobordism invariant of $\Omega_1^{\Spin}=\Z_2$. 
{The 1d manifold generator of $\tilde{\eta}$ is a circle $S^1$ with a periodic boundary condition (i.e., Ramond) for the fermion.}\\
$\bullet$  The Arf invariant \cite{Arf1941} is a mod 2 cobordism invariant of $\Omega_2^{\Spin}=\Z_2$,
 whose quantum matter realization is the 1+1d Kitaev fermionic chain \cite{Kitaev2001chain0010440} {whose each open end hosts a 0+1d Majorana zero mode}.\\
$\bullet$ The $(\CA_{{\Z_2}})^5$ is a mod 2 class purely bosonic topological invariant, which corresponds to a 5d bosonic SPT phase given by
 the group cohomology class data $\H^5(\B\Z_2,\U(1))=\Z_2$, which is also one of the $\Z_2$ generators in $\Omega_5^{\SO}(\B\Z_2)$.
 }

\begin{enumerate}[leftmargin=-1.mm]
\item
When ${\upnu}$ is odd, such as ${\upnu}=1,3,5,7,\dots  \in \Z_{16}$, \\
\Refe{Hsieh2018ifc1808.02881} suggested that the symmetry-extension method \cite{Wang2017locWWW1705.06728} cannot construct a symmetric gapped TQFT.  
Furthermore, Cordova-Ohmori \cite{Cordova1912.13069} proves that a symmetry-preserving gapped TQFT phase is impossible for
this odd ${\upnu} \in \Z_{16}$ anomaly from $\Omega_5^{\Spin \times_{\Z_2} \Z_4}=\Z_{16}$.
The general statement in \cite{Cordova1912.13069} is that 
given an anomaly index $\upnu \in \Omega_5^{\Spin \times_{\Z_2} \Z_{4}}=\Z_{16}$,
we can at most construct a fully symmetric gapped TQFT 
\emph{if and only if} 
$ 4 \mid 2 \upnu.$
Namely, 4 has to be a divisor of $2 \upnu$.
Apparently, the $4 \mid 2 \upnu$ is true only when $\upnu$ is even.

Since ${\upnu=-N_{\text{generation}}}$, the case of ${\upnu}=1$ (for a single generation) and ${\upnu}=3$ (for three generations) are particularly important 
for the high energy physics phenomenology.
{This means that we are not able to
directly construct any 4d symmetric gapped TQFT that explicitly matches the same $\Z_{16}$ anomaly 
for one right-handed neutrino (${\upnu}=1$) or three right-handed neutrinos (${\upnu}=3$) alone.}

\item When ${\upnu}$ is even, such as ${\upnu}=2,4,6,8,\dots  \in \Z_{16}$, 
\Refe{Hsieh2018ifc1808.02881, Wan2019sooWWZHAHSII1912.13504, PrakashJW2011.13921, PTWtoappear} suggested that the symmetry-extension method \cite{Wang2017locWWW1705.06728} can trivialize the 't Hooft anomaly. 
Furthermore, Cordova-Ohmori \cite{Cordova1912.13069} shows that there is no obstruction to construct 
a symmetry-preserving gapped TQFT phase for any even ${\upnu_{\text{even}} } \in \Z_{16}$.
We can verify the claim by rewriting \eq{eq:5dSPT-4dTQFT} in terms of the $(\frac{\upnu_{\text{even}}}{2}) \in \Z_{8}$ index:
\begin{multline} \label{eq:5dSPT-4dTQFT-Z8}
{{\bf Z}_{{\text{$5$d-iTQFT}}}^{(\upnu_{\text{even}}=2)}}
=
\exp(\frac{2\pi \ii}{16} \cdot{\upnu_{\text{even}}} \cdot \eta(\text{PD}(\CA_{{\Z_2}})) \bigg\rvert_{M^5})
= 
\exp(\frac{2\pi \ii}{8} \cdot (\frac{{\upnu_{\text{even}}}}{2}) \cdot  \Big( 
\text{ABK} (\text{PD}((\CA_{{\Z_2}})^3) ) 
\Big) \bigg\rvert_{M^5}) \\
= 
\exp(\frac{2\pi \ii}{8} \cdot (\frac{{\upnu_{\text{even}}}}{2}) \cdot  \Big(  4 \cdot \text{Arf} (\text{PD}((\CA_{{\Z_2}})^3) ) 
+ 2 \cdot{\tilde{\eta}} (\text{PD}((\CA_{{\Z_2}})^4) )   + (\CA_{{\Z_2}})^5 \Big) \bigg\rvert_{M^5}),
\end{multline}
with a 2d Arf-Brown-Kervaire (ABK) invariant {which is also known as the Pin$^{-}$-structure $\Z_8$-class of
iTQFT of the 1+1d Fidkowski-Kitaev fermionic chain \cite{FidkowskifSPT1, FidkowskifSPT2} 
with a time reversal $T^2=+1$ symmetry.}
{Notice that \eq{eq:5dSPT-4dTQFT-Z8} can become trivialized if we can trivialize the $(\CA_{{\Z_2}})^3$ factor.
In fact, the 
$(\CA_{{\Z_2}})^3$ can be trivialized by the symmetry extension \cite{Wang2017locWWW1705.06728}, written
in terms of the group extension of a short exact sequence:
$$
0  \to \Z_2 \to {\Z_{4,X}} \to \frac{\Z_{4,X}}{\Z_2^F} \to 0.
$$
Namely, the 2-cocycle topological term $(\CA_{{\Z_2}})^3$ in $\H^3(\B(\frac{\Z_{4,X}}{\Z_2^F}),\U(1))$ becomes a coboundary once we
lifting the $\frac{\Z_{4,X}}{\Z_2^F}$-gauge field $\CA_{{\Z_2}}$ 
to a ${\Z_{4,X}}$ gauge field in
$\H^3(\B{\Z_{4,X}},\U(1))$.} 
So this suggests that the following symmetry extension for the spacetime-internal symmetry, written
in terms of the group extension of a short exact sequence:\footnote{See more discussions
in Sec.~5 of \cite{JW2006.16996}, and in \cite{PTWtoappear}.}
\bea \label{eq:Z2-SpinxZ4}
1  \to [\Z_2] \to \Spin \times {\Z_{4,X}} \to \Spin \times_{\Z_2^F} {\Z_{4,X}} \to 1.
\eea
can fully trivialize any even ${\upnu_{\text{even}}} \in \Z_{16}$ cobordism invariant given in \eq{eq:5dSPT-4dTQFT-Z8}.
The $[\Z_2]$ means that we can gauge the \emph{anomaly-free} 
normal subgroup $[\Z_2]$ in the total group $\Spin \times {\Z_{4,X}}$.
This symmetry extension \eq{eq:Z2-SpinxZ4} also means that a 4d $[\Z_2]$ gauge theory preserves the 
$\Spin \times_{\Z_2^F} {\Z_{4,X}}$ symmetry while 
 also saturates the even ${\upnu_{\text{even}}} \in \Z_{16}$ anomaly.
This 4d $[\Z_2]$ gauge theory is the \emph{anomalous} symmetric gapped non-invertible TQFT 
(with 't Hooft anomaly of $\Spin \times_{\Z_2^F} {\Z_{4,X}}$-symmetry)
desired in the \cred{Scenario \ref{Z4XTQFT}}.

\end{enumerate}


Since a symmetric anomalous 4d TQFT only exists with even ${\upnu_{\text{even}}} \in \Z_{16}$,
below we formulate the path integral ${\bf Z}_{{\text{4d-TQFT}}}^{(\upnu_{\rm 4d}=2)}[\CA_{{\Z_4}}]$
of the root phase $\upnu_{\rm 4d}=2$.
We generalize the boundary TQFT construction in the Section 8 of \cite{GuoJW1812.11959}.
With
${\upnu_{\text{even}}} =2 \in \Z_{16}$,
we have \eq{eq:5dSPT-4dTQFT-0} with the input of 5d bulk iTQFT \eq{eq:5dSPT-4dTQFT-Z8},
then we can explicitly construct the partition function on a 5d manifold $M^5$ with a 4d boundary 
$M^4 \equiv \partial M^5$ as,\footnote{We use the $\smile$ notation for the cup product between cohomology classes,
or between a cohomology class and a fermionic topological invariant (paired via a Poincar\'e dual PD).
We often make the cup product $\smile$ and the Poincar\'e dual PD implicit.
We use the $\cup$ notation for the surgery gluing the boundaries of two manifolds within relative homology classes.
So the $M_1 \cup M_2$ means gluing the boundary $\partial M_1 = \overline{\partial M_2}$ such that the common orientation of
$\overline{\partial M_2}$ is the reverse of ${\partial M_2}$.
}
{
\begin{multline} \label{eq:5dSPT-4dTQFT-explicit}
{\bf Z}_{{\text{5d-iTQFT}}}^{}[\CA_{{\Z_4}}]
\cdot 
{\bf Z}_{{\text{4d-TQFT}}}^{}[\CA_{{\Z_4}}]
=
 \sum_{c\in {\partial'}^{-1}(\partial [\text{PD}(\CA^3)])}\e^{\frac{2\pi\ii}{8}\text{ABK}(c\cup \text{PD}(\CA^3))} \quad\quad\quad\quad\quad\quad\\
\cdot \frac{1}{2^{|\pi_0(M^4)|}} \sum_{\substack{a\in C^1(M^4,\Z_2) ,\\b\in C^2(M^4,\Z_2)} }(-1)^{\int_{ M^4} a (\delta b+\CA^3)} \cdot \e^{\frac{2\pi\ii}{8}\text{ABK}(c\cup \text{PD}'( b ))}
.
\end{multline}} 
We write the mod 2 cohomology class $\Z_2$ gauge field as 
{$\CA \equiv \CA_{{\Z_2}}  \in \H^1(M^5, \frac{\Z_{4,X}}{\Z_2^F})$,
while the $\Spin \times_{\Z_2^F} \Z_{4,X}$ gauge field $\CA_{{\Z_4}}$
is not an ordinary abelian discrete gauge field but with the extra constraint $w_2(TM) = \CA_{{\Z_2}}^2 = \CA^2$}.
Here come some remarks on this 5d bulk iTQFT-4d boundary TQFT partition function \eq{eq:5dSPT-4dTQFT-explicit}:
\begin{enumerate} [leftmargin=.mm, label=\textcolor{blue}{\arabic*)}., ref={\arabic*)}]
\item 
The $(\text{PD}(\CA^3))$ is a 2d manifold taking the Poincar\'e dual (PD) of 3-cocycle $\CA^3$ in the $M^5$; 
but the 2d manifold $(\text{PD}(\CA^3))$ may touch the 
the 4d boundary $\partial M^5 =M^4$.
The 1d boundary $\partial (\text{PD}(\CA^3))$ can be regarded as the 1d intersection 
between the 2d $(\text{PD}(\CA^3))$ and the $M^4$.

\item \noindent 
More precisely, for a $\Spin \times_{\Z_2} \Z_4$ manifold $M^5$ with a boundary, 
we have used the Poincar\'e-Lefschetz duality for a manifold with boundaries:
\bea
{\CA^3} \in \H^3(M^5,\Z_2)\xrightarrow{\cong} \H_2(M^5, M^4,\Z_2)\ni \text{PD}(\CA^3).
\eea

\item \noindent
For any pair $(\rS,\rS')$, where $\rS'$ is a subspace of $\rS$, the short exact sequence of chain complexes 
\bea
0\to C_*(\rS')\to C_*(\rS)\to C_*(\rS,\rS')\to 0,
\eea
with $C_n(\rS,\rS') \equiv C_n(\rS)/C_n(\rS')$, induces a long exact sequence of homology groups
\bea
\cdots\to \H_n(\rS')\to \H_n(\rS)\to \H_n(\rS,\rS')\xrightarrow{\partial} \H_{n-1}(\rS')\to\cdots.
\eea
Here $\H_n(\rS,\rS')$ is the relative homology group, and $\partial$ is the boundary map.\\
$\bullet$ Take $(\rS,\rS')=(\text{PD}(\CA^3),\partial \text{PD}(\CA^3))$, we denote the boundary map by $\partial$: 
\bea \label{eq:map-partial}
\H_2( \text{PD}(\CA^3), \partial \text{PD}(\CA^3) ) \xrightarrow{\partial} \H_{1}( \partial \text{PD}(\CA^3)).
\eea
Here $\text{PD}(\CA^3)$ is not a closed 2-manifold, but which has a boundary closed 1-manifold $\partial \text{PD}(\CA^3)$. \\
$\bullet$  Take $(\rS,\rS')=(M^4= \partial M^5, \partial \text{PD}(\CA^3))$, we denote another boundary map by $\partial_1$:
\bea \label{eq:map-partial1}
\H_2(M^4, \partial \text{PD}(\CA^3) ) \xrightarrow{\partial_1} \H_{1}( \partial \text{PD}(\CA^3)).
\eea 
Both $M^4= \partial M^5$ and $\partial \text{PD}(\CA^3)$ are closed manifolds, of 4d and 1d, respectively.

\item \noindent
Now, the $c$ is defined as a 2d surface living on the boundary $M^4$.
The ${\partial_1}c$ uses the boundary map \eq{eq:map-partial1}'s $\partial_1$ of $c$ on the $M^4$.
We can compensate the 2-surface $\text{PD}(\CA^3)$ potentially with a 1-boundary, 
by gluing it with $c$  to make a closed 2-surface.
To do so, we require both $\text{PD}(\CA^3)$ and $c$ share the same 1d boundary.

\item 
\noindent
 The ${c\in {\partial_1}^{-1}(\partial [\text{PD}(\CA^3)])}$ also means
\bea \label{eq:partial'c=partialPDA3}
{{\partial_1} c = \partial [\text{PD}(\CA^3)]  = [\partial \text{PD}(\CA^3)]}.
\eea
\noindent
$\bullet$ The $[\text{PD}(\CA^3)]$ means the fundamental class and the relative homology class of the 2d manifold $\text{PD}(\CA^3)$.\\
 \noindent
$\bullet$ The {$\partial [\text{PD}(\CA^3)] = [\partial \text{PD}(\CA^3)]$} means the boundary of the fundamental class
(via the boundary $\partial$ map in \eq{eq:map-partial})
is equivalent to the fundamental class of the boundary $\partial$ of $\text{PD}(\CA^3)$.
Beware that the two $\partial$ operations in {$\partial [\text{PD}(\CA^3)] = [\partial \text{PD}(\CA^3)]$} have different meanings. \\
$\bullet$ \Eqn{eq:partial'c=partialPDA3} exactly matches the requirement that the bulk 2-surface $\text{PD}(\CA^3)$ (living in $M^5$) and
the boundary 2-surface $c$ (living on $M^4$) share the same 1d boundary.

\item 
\noindent Note that when $M^5$ is a closed manifold with no boundary $M^4= \partial M^5= \emptyset$ thus $c = \emptyset$,
then the term 
$$\sum_{c\in {\partial_1}^{-1}(\partial [\text{PD}(\CA^3)])}\e^{\frac{2\pi\ii}{8}\text{ABK}(c\cup \text{PD}(\CA^3))}
\text{ is equivalently reduced to }
\e^{\frac{2\pi\ii}{8}\text{ABK}( \text{PD}(\CA^3))}
={{\bf Z}_{{\text{$5$d-iTQFT}}}^{(\upnu_{\text{even}}=2)}}.$$
This is a satisfactory consistent check, consistent with the 5d bulk-only iTQFT at ${\upnu_{\text{even}}}=2$ in \eq{eq:5dSPT-4dTQFT-Z8}.

\item \noindent
The 
$a\in C^1(M^4,\Z_2)$ means that $a$ is a 1-cochain,
and the $b\in C^2(M^4,\Z_2)$ means that {$b$ is a 2-cochain}.
The factor {$(-1)^{\int_{ M^4} {a (\delta b}+\CA^3)}= 
\exp(\ii \pi {\int_{ M^4} {a (\delta b}+\CA^3)})$} gives the weight of the 
4d $\Z_2$ gauge theory. 
The {$a \delta b$} term is the level-2 BF theory written in the mod 2 class.\footnote{The continuum QFT version of this $\Z_2$ gauge theory is 
$\exp(\ii  {\int_{ M^4} { \frac{2}{2 \pi}  a \dd b +\frac{1}{\pi^3} a \CA^3} )})$,
where
the $b$ integration over a closed 2-cycle, $\oiint b$, can be $n \pi$ with some integer $n \in \Z$.
The $a$ and $\CA$ integration over a closed 1-cycle, $\oint a$ and $\oint \CA$, can be $n \pi$ with some integer $n \in \Z$.\\
{Another alternative possibility of 4d TQFT of \Eq{eq:5dSPT-4dTQFT-explicit} can be
$
\frac{1}{2^{|\pi_0(M^4)|}} \sum_{\substack{a\in C^1(M^4,\Z_2) ,\\b\in C^2(M^4,\Z_2)} }(-1)^{\int_{ M^4} b (\delta a+\CA^2)} \cdot \e^{\frac{2\pi\ii}{8}\text{ABK}(c\cup \text{PD}'( \CA a ))}
$. 
{In contrast to \eq{eq:trivializationA2}, this 4d TQFT is constructed out of a different trivialization 
$w_2(TM) = \CA^2 = \delta a =0$ under the pullback \eq{eq:Z2-SpinxZ4} to a $\Spin \times_{} {\Z_{4,X}}$-structure.}
The continuum version of this $\Z_2$ gauge theory has a different expression
as $\exp(\ii  {\int_{ M^4}\frac{2}{2 \pi} b \dd a+ \frac{1}{\pi^2} b \CA^2)})$.
{ 
Either 4d $\Z_2$ gauge theory sits at the normal subgroup $ [\Z_2]$ of the group extension \eq{eq:Z2-SpinxZ4}.
Although the $ [\Z_2]$ is abelian, this 4d TQFT 
actually exhibits \emph{non-abelian topological order} due to the fermionic nature of $\Spin \times_{\Z_2^F} {\Z_{4,X}}$ and the fermionic invariant ABK.\\
The non-abelian nature of this 4d TQFT is similar to the non-abelian nature of 3d $\Z_2$ gauge theory obtained from gauging 
the $\Z_2$-onsite symmetry of the odd class of 2+1d  fermionic topological superconductor from
the $\Omega_3^{\Spin \times \Z_2} =\Z_8$ classification (e.g., Section 8 of \cite{Putrov2016qdo1612.09298PWY}, and \cite{GuoJW1812.11959}). 
The braiding and fusions statistics of vortices of these TQFTs are \emph{non-abelian}.
Moreover,
we may require \emph{additional symmetry extension} beyond \eq{eq:Z2-SpinxZ4} to construct generic symmetric 4d TQFTs.} 
}
}
The $\delta b$ is a coboundary operator $\delta$ acting on $b$.
The path integral sums over these distinct cochain classes.

The variation of {$a$} gives the equation of motion {$(\delta b+\CA^3) = 0 \mod 2.$}
In the path integral, we can integrate out {$a$} to give the same constraint {$(\delta b+\CA^3) = 0 \mod 2.$}
This is precisely the trivialization of the second cohomology class,
\bea \label{eq:trivializationA2}
{ \CA^3= (\CA_{{\Z_2}})^3 =\delta b \mod 2} 
\eea
so the {3-cocycle becomes a 3-coboundary which splits to a 2-cochain $b$.} 
This exactly matches the condition imposed by the symmetry extension \eq{eq:Z2-SpinxZ4}:
$1  \to [\Z_2] \to \Spin \times {\Z_{4,X}} \to \Spin \times_{\Z_2^F} {\Z_{4,X}} \to 1$,
where the {$\CA^3$} term in $ \Spin \times_{\Z_2^F} {\Z_{4,X}}$ becomes trivialized as a coboundary 
{$\CA^3 =\delta b$} 
(so {$\CA^3= 0$} in terms of a cohomology or cocycle class)
in $\Spin \times {\Z_{4,X}}$.

\item \noindent
The $\frac{1}{2^{|\pi_0(M^4)|}}$ factor mod out the gauge redundancy for the boundary 4d $\Z_2$ gauge theory. 
The $\pi_0(M^4)$ is the zeroth homotopy group of $M^4$, namely
the set of {all path components} of $M^4$.
Thus, we only sum over the {gauge equivalent} classes in the path integral.

\item \noindent
The ${\ABK}(c\cup \text{PD}(\CA^3))$ is defined on a 2d manifold with $\Pin^-$ structure.
Recall the $M^5$ has the $\Spin \times_{\Z_2} {\Z_{4}}$ structure.
If $M^5$ is closed, then there is a natural Smith map to induce the 2d $\Pin^-$ structure on the closed surface via $\text{PD}(\CA^3)$.
However, the $M^5$ has a boundary $M^4$, so $ \text{PD}(\CA^3)$ may not be closed  --- the previously constructed closed 2-surface $(c\cup \text{PD}(\CA^3))$
is meant to induce a 2d $\Pin^-$ structure.\footnote{We do not yet know whether it is always possible to induce a unique 2d $\Pin^-$ structure on
$(c\cup \text{PD}(\CA^3))$ for any possible pair of data $(M^5,M^4= \partial M^5)$ given {any $M^5$ with $\Spin \times_{\Z_2} {\Z_{4}}$ structure}. However, we claim that it is possible
to find some suitable $M^4$ so that the 2d $(c\cup \text{PD}(\CA^3))$ has $\Pin^-$ induced, thus in this sense the ${\ABK}(c\cup \text{PD}(\CA^3))$ is defined.
For physics purposes, it is enough that we can {firstly} focus on studying the theory on these types of $(M^5,M^4= \partial M^5)$.
\label{ft:inducePin-1}} 
Then we compute the ABK on this closed 2-surface $(c\cup \text{PD}(\CA^3))$.

\item \noindent
Let us explain 
the other term ${\text{ABK}(c\cup \text{PD}'(\CA a))}$ of the 4d boundary TQFT in \eq{eq:5dSPT-4dTQFT-explicit}.
The $\text{PD}'$ is the Poincar\'e dual on $M^4= \partial M^5$.
Since 
the fundamental classes of $M^5$ and $M^4= \partial M^5$ are related by
\bea
[M^5]\in \H_5(M^5, M^4,\Z_2)\xrightarrow{\partial}\H_4( M^4,\Z_2)\ni[ M^4]  \equiv [\partial M^5].
\eea
Here we have the following relations:\footnote{{Let us clarify
the notations: $\partial$, $\partial'$, and $\partial_1$. 
The boundary notation $\partial$ may mean as (1) taking the boundary, or (2) in the boundary map of relative homology class in \eq{eq:map-partial}. 
It should be also clear to the readers 
that\\ 
$\bullet$ the $\partial$ is associated with the operations on objects living in the bulk $M^5$ or ending on the boundary $M^4$,\\
$\bullet$ while the $\partial'$ is associated with the operations on objects living on the boundary $M^4$ alone.\\
$\bullet$ The $\partial_1$ is defined as another boundary map in \eq{eq:map-partial1}.}}
\bea \label{eq:PD-cap-relation}
\text{PD}&=&[M^5]\cap, \cr
\text{PD}' &=&[ M^4]\cap=[\partial M^5]\cap =\partial[ M^5]\cap,\\
{{\partial'} \text{PD}'(b)} &=&
{{\text{PD}'(\delta b )=\text{PD}'(\CA^3)=\partial [\text{PD}(\CA^3)]=\partial_1 c.}}
\eea
$\bullet$ The cap product $\cap$ here is to define PD homology class, such that $\text{PD}(\CA) = [M^5] \cap \CA$.\\
$\bullet$  Here we use $[M^4]=[\partial M^5]=\partial [M^5] $:
the fundamental class of boundary of $M^5$ gives
the boundary of fundamental class.\\
$\bullet$ {The \eq{eq:PD-cap-relation}'s first equality {${\partial'} \text{PD}'(b )  =\text{PD}'(\delta  b)$} uses
the coboundary operator $\delta$ on the cohomology class {$b$}.}\\
$\bullet$ The \eq{eq:PD-cap-relation}'s second equality {$\text{PD}'(\delta b)=\text{PD}'(\CA^3)$} 
uses the condition $\delta \CA =0$ and the trivialization condition \eq{eq:trivializationA2}: {$\delta b = \cA^3$}. \\
$\bullet$ The \eq{eq:PD-cap-relation}'s third equality $\text{PD}'(\CA^3)=\partial [\text{PD}(\CA^3)]$,
we use ``\emph{the naturality of the cap product}''. 
There are natural pushforward
and pullback maps on homology and cohomology, 
related by the projection formula, also known as ``the naturality of the cap product.''
\\
$\bullet$ The \eq{eq:PD-cap-relation}'s last equality $\partial [\text{PD}(\CA^3)]=\partial_1 c$ is based on \eq{eq:partial'c=partialPDA3}. 
Importantly, as a satisfactory consistency check, this also shows that the 2-surface $c$ obeys: 
\bea
\text{the $c$
in  ${\ABK}(c\cup \text{PD}(\CA^3))$ is the same $c$ in ${\text{ABK}(c\cup \text{PD}'({b}))}$.}
\eea
$\bullet$ As before, the union $(c\cup\text{PD}'({b}))$ is a closed 2-surface and we induce a $\Pin^-$ structure on this 2-surface.
So we can compute the ABK on this closed 2-surface $(c\cup\text{PD}'({b}))$ on the $M^4$.\footnote{Similar to Footnote \ref{ft:inducePin-1},
 we can find some suitable $M^4$ so that the 2d $\Pin^-$ is induced, thus in this sense the ${\ABK}(c\cup\text{PD}'(\cred{b}))$ is defined.
\label{ft:inducePin-2} }

\end{enumerate}

In summary, we have constructed the 4d 
fermionic discrete gauge theory in \eq{eq:5dSPT-4dTQFT-explicit}
preserving the $(\Spin \times_{\Z_2^F} {\Z_{4,X}})$-structure, namely it is a $(\Spin \times_{\Z_2^F} {\Z_{4,X}})$-{symmetric} TQFT
but with ${\upnu_{\text{even}}}=2 \in \Z_{16}$ anomaly.
This can be used to compensate the anomaly 
$\upnu =- N_{\text{generation}}\mod 16$
with $N_{\text{generation}}=2$, two generations of missing right-handed neutrinos.
We could not however directly construct the symmetric gapped TQFT for 
$\upnu$ is odd (thus symmetric TQFTs not possible for $N_{\text{generation}}=1$ or $3$), due to the obstruction found in 
\cite{Hsieh2018ifc1808.02881, Cordova1912.13069}.

\subsection{General Principle}

The discussion in \Sec{sec:sym-ext} says that only for 
the even integer ${\upnu_{4d,\text{even}}} \in \Z_{16}$ does the $(\Spin \times_{\Z_2^F} {\Z_{4,X}})$-symmetry-preserving TQFT exist.  
This prompts us to improve
the SM version \eq{eq:ZSMtotal} into:
\bea \label{eq:ZSMtotaleven}
\bZ_{\UU} [\CA_{{\Z_4}}]
\equiv {{\bf Z}_{\substack{\text{5d-iTQFT/}\\ \text{4d-SM+TQFT}}}[\CA_{{\Z_4}}]
\equiv
{\bf Z}_{{\text{5d-iTQFT}}}^{(-\upnu_{\rm 5d})}[\CA_{{\Z_4}}]
\cdot 
{\bf Z}_{{\text{4d-TQFT}}}^{(\upnu_{\rm 4d,even})}[\CA_{{\Z_4}}]\cdot 
\bZ_{\SM}^{(n_{\nu_{e,R}}, n_{\nu_{\mu,R}}, n_{\nu_{\tau,R}})}
[\CA_{{\Z_4}}]
.}
\eea
The GUT version \eq{eq:Zsu5total} should be adjusted into:
\bea \label{eq:Zsu5totaleven}
{\bZ_{\UU} [\CA_{{\Z_4}}]
\equiv
{\bf Z}_{\substack{\text{5d-iTQFT/}\\ \text{4d-GUT+TQFT}}}[\CA_{{\Z_4}}]
\equiv
{\bf Z}_{{\text{5d-iTQFT}}}^{(-\upnu_{\rm 5d})}[\CA_{{\Z_4}}]
\cdot 
{\bf Z}_{{\text{4d-TQFT}}}^{(\upnu_{\rm 4d,even})}[\CA_{{\Z_4}}]
\cdot 
\bZ_{\GUT}^{(n_{\nu_{e,R}}, n_{\nu_{\mu,R}}, n_{\nu_{\tau,R}})}
[\CA_{{\Z_4}}].
}
\eea
Also the anomaly constraint \eq{eq:anomaly-match} becomes:
\bea
{
  (-{(N_{\text{gen}}=3)}+ n_{\nu_{e,R}} + n_{\nu_{\mu,R}} + n_{\nu_{\tau,R}} + \upnu_{\rm 4d,even} - \upnu_{\rm 5d} ) = 0 \mod 16.
}\eea
This implies that the existence of symmetry-preserving 4d TQFT sector requires the following:\footnote{The nonperturbative global anomaly {cancellation} 
constraint $(-{(N_{\text{gen}}=3)}+ n_{\nu_{e,R}} + n_{\nu_{\mu,R}} + n_{\nu_{\tau,R}} + \upnu_{\rm 4d,even} - \upnu_{\rm 5d}) = 0 \mod 16$ provides
the capacity for many kinds of the Ultra Unification model building. 
For example, the 4d $\Z_{4,X}$-symmetry preserving TQFT sector can take the index $\upnu_{\rm 4d,even}=0,2,4,\dots$ for any even integer. 
There are many (perhaps countably infinite) types of TQFTs for each index $\upnu_{\rm 4d,even}$. 
But to be more economic, we can ask for the minimum degrees of freedom required by a TQFT for any given index $\upnu_{\rm 4d,even}$.
Also for HEP phenomenological purposes, by taking account of the experimentally observed neutrino mass eigenstates splitting: \\
$\bullet$ if one use the \emph{conventional quadratic mass mechanism} to generate the observed neutrino masses,
one may propose to have 
at least two generations of right-handed neutrinos, which means that a possible phenomenological input $n_{\nu_{e,R}} + n_{\nu_{\mu,R}} + n_{\nu_{\tau,R}} \geq 2$.
A viable Ultra Unification candidate can be, for example,  $n_{\nu_{e,R}} + n_{\nu_{\mu,R}} + n_{\nu_{\tau,R}} = 2$, $ \upnu_{\rm 4d,even}=2$, and $\upnu_{\rm 5d}=1$,
which saturates the anomaly {cancellation} 
and some phenomenological constraints.\\
$\bullet$ if we use the interacting topological mass $\Delta_{\TQFT}$ and its topological defect energy subgap $\Delta_{\rm sub}$ 
to account for the observed neutrino masses (see \Sec{label:Summary} and \Fig{fig:Fig9} for illustrations), 
then we may have different choices of the number of right-handed neutrinos, written as $( \sum_{j=e,\mu,\tau,\dots }n_{\nu_{j,R}})$ in \eq{eq:anomaly}.\\
So far, we mainly use the cobordism theory to study the invertible anomalies and invertible topological field theories, 
and we also use the cohomology data to construct non-invertible topological quantum field theories. 
However, once the discrete symmetries (such as $\Z_{4,X}$) are dynamically gauged, it is more natural 
to use the mathematical category or higher category theories to characterize the Topological Phase Sectors.}
$$
{\text{$(n_{\nu_{e,R}} + n_{\nu_{\mu,R}} + n_{\nu_{\tau,R}} - \upnu_{\rm 5d})$ must be an odd integer.}}
$$

Now we have derived an Ultra Unification path integral in \eq{eq:ZSMtotaleven} and \eq{eq:Zsu5totaleven},
including
4d SM \eq{eq:ZSM},
4d GUT \eq{eq:Zsu5},
5d iTQFT \eq{eq:5dSPT-4dTQFT} and
4d TQFT \eq{eq:5dSPT-4dTQFT-explicit},
comprising many Scenarios and their linear combinations enlisted in \Sec{sec:Consequences}: \ref{Massless}, \ref{Dirac}, \ref{Majorana}, \ref{Z4XTQFT}, and \ref{Z45dSPT}.
Then we can dynamically gauge the appropriate bulk-boundary global symmetries,
promoting the theory to a bulk gauge theory in Scenarios \ref{Z45dSET} and \ref{Z45dSET2},
or break some of the (global or gauge) symmetries to \ref{Z4XbreakTQFT}. 

\paragraph{General Principle\\}
In summary, we propose a general principle behind the Ultra Unification:\\ 
1. We start with a QFT in general as an effective field theory (EFT) given some full 
{spacetime-internal symmetry} $G$, say in a D dimensional (Dd) spacetime.\\
2. We check the anomaly and cobordism constraint given by $G$ via 
computing 
$\Omega^{\text{D+1}}_{G} \equiv \TP_{\text{D+1}}(G)$.\\
3. We check the {anomaly index} of the Dd QFT/EFT constrained by cobordism $\Omega^{\text{D+1}}_{G} \equiv \TP_{\text{D+1}}(G)$.\\
4. If all anomalies are {canceled}, then we do not require any new hidden sector to define Dd QFT/EFT.\\ 
5. If some anomalies are {not canceled}, we have two perspectives, \emph{either} regarding the $G$-symmetry as a global symmetry with 't Hooft anomaly at a lower energy;
\emph{or} regarding the $G$-symmetry and its anomaly still persist at a higher energy 
(thus we can further assume the $G$ symmetry is dynamically gauged at a higher energy due to ``no global symmetry in quantum gravity reasonings''), 
then either\\ 
(1) we need to break some symmetry out of $G$ to eliminate the anomaly, or\\ 
(2) we extend the symmetry $G$ to an appropriate $\tilde{G}$ to trivialize nonperturbative global anomalies
in $\Omega^{\text{D+1}}_{\tilde{G}} \equiv \TP_{\text{D+1}}(\tilde{G})$, or \\
(3) we propose new hidden sectors appending to Dd QFT/EFT, with a schematic path integral
(say if we add Dd-{TQFT or CFT} and {(D+1)d-iTQFT} onto the original theory):
\bea \label{eq:ZSMtotal-new}
\boxed{{\bf Z}_{\substack{\text{(D+1)d-iTQFT/}\\ \text{Dd-QFT/EFT+{TQFT or CFT}}}}[\CA_{}]
\equiv
{\bf Z}_{{\text{(D+1)d-iTQFT}}}^{}[\CA_{}]
\cdot 
{\bf Z}_{{\text{Dd-{TQFT or CFT}}}}^{}[\CA_{}]\cdot 
\bZ_{\text{Dd-QFT/EFT}}^{}
[\CA_{}]
.}
\eea
Here $\CA$ is a generic $G$-symmetry background field, that is also to be dynamically gauged at a higher energy.

\subsection{Detect Topological {Phase} Sectors and the Essence of Ultra Unification}

\paragraph{Detect Topological Forces
\\}
By looking at \Fig{fig:BSM-UU-tree}, in the Higgs vacuum where our Standard Model effective field theory (SM EFT) resides in,
we have only detected the Strong, Electromagnetic, and Weak in the subatomic physics. 
The GUT forces are weaker than the Weak force, and the Topological Force is further weaker than the GUT and Weak forces.
So how could we experimentally detect Topological Forces?

Notice that the gravity is further weaker than all other forces. 
(So how could we experimentally detect gravity?)
But the gravity has accumulative effects that only have the gravitational attractions.
Without doubt, the gravity has been detected by everyone and by all astrophysics and cosmology observations. 
The gravity had been detected first in the human history among all the forces!

Similarly, although Topological Force is also weak (but stronger than the gravity),
Topological Force is infinite-range or long-range which does \emph{not decay in the long distance}, and mediates between
the linked worldline/worldsheet/worldvolume trajectories of the charged (point-like or extended) objects
via fractional or categorical anyonic statistical interactions.
So in principle, \emph{we may have already experienced Topological Force in our daily life, in a previously 
scientifically unnoticed way.}\footnote{
%
{For example, if the $\Z_{4,X}$ is dynamically gauged, there is a dynamical discrete gauge Wilson line connecting all SM fermions living in 4d SM or GUT (e.g., quarks and electrons in our body).
Namely, the SM fermions can live at the open ends of the $\Z_{4,X}$ gauged Wilson line.
Thus there could be long-distance topological interactions and communications between the $\Z_{4,X}$-gauge charged objects. \\[2mm] 
Moreover, the $\Z_{4,X}$ gauged Wilson line as a $\cA_{\Z_4}$ gauge field on the 4d theory can be leaked  
into {the 5d bulk theory as a $\cA_{\Z_2}$ gauge field. In the 5d bulk, a nontrivial link configuration can 
be charged under the other end of  $\cA_{\Z_2}$ \cite{JW2006.16996}.}\\[2mm] 
%
{It is worthwhile to emphasize that the new sectors that we propose (the 4d gapped phase with TQFT, the 4d gapless phase with CFT,
and the 5d gapped phase with iTQFT or TQFT) are \emph{invisible} to SM gauge forces ($su(3) \times su(2) \times u(1)$)
also \emph{invisible} to the $su(5)$ GUT forces. However, 
these new sectors are \emph{detectable} via Topological Forces (i.e., statistical interaction via discrete gauge forces).
These new sectors are also \emph{detectable} via the gravitations. 
}\\[2mm]
In a metaphysics sense, perhaps the long-distance Topological Force might be related to some of the unexplained mysterious 
phenomena. In any case, every phenomenon and every law of Nature should be explained by mathematics and physics principles.} \\[1mm] }

\paragraph{Neutrino Oscillations and Dark Matter\\}

In fact, \Refe{JW2006.16996} had proposed that the Topological Force may {cause (thus be detected by) the phenomena} of neutrino oscillations.
We can consider the Majorana zero modes of the vortices in the 4d TQFT defects. 
The left-handed neutrinos ({confirmed by} the experiments) are nearly gapless/massless. 
When the left-handed  neutrinos traveling through the 4d TQFT defects,
 we may observe nearly {gapless} neutrino flavor oscillations interfering with the Majorana zero modes trapped by the vortices/vortex strings/monopoles in the 4d TQFT defects (see
\Sec{label:Summary} and \Fig{fig:Fig9} for illustrations).
On the other hand, the gapped heavy excitations (point or extended objects) 
of Topological Phase Sector may be a significant contribution to Dark Matter \cite{toappear}.

If the TQFT energy gap $\Delta_{\rm TQFT}$ is large, the heavy excitations above the $\Delta_{\rm TQFT}$ may contribute the {\bf\emph{heavy Dark Matter}}.
In contrast, if the $\Delta_{\rm TQFT}$ is small compared to the SM's particle masses, 
or if the new hidden sector contains CFT, then the new sector contributes the {\bf\emph{light Dark Matter}}.

\paragraph{The Essence of Ultra Unification\\}

Finally, we come to the essence of {\bf Ultra Unification}. What is unified after all?
We have united the {\bf Strong, Electromagnetic, Weak, GUT forces, and Topological forces into the same theory}
in the Ultra Unification QFT/TQFT path integral 
(that this theory can also be coupled to the curved spacetime geometry and gravity, at least well-defined in a background non-dynamical way).
See \Fig{fig:Fig2SM}, \Fig{fig:Fig3GUT}, \Fig{fig:Fig4extra}, and \Fig{fig:Fig5UU}  for illustrations.

However, the Grand Unification 
\cite{Georgi1974syUnityofAllElementaryParticleForces, Fritzsch1974nnMinkowskiUnifiedInteractionsofLeptonsandHadrons}
united
the three gauge interactions of the SM into a single electronuclear force
under a simple Lie group gauge theory.
Do we have any equivalent statement to also unite {\bf Strong, Electromagnetic, Weak, GUT forces and Topological forces into a single force} at a high enough energy?
We believe that the definite answer 
relies on studying the details of analogous topological quantum phase transitions \cite{JW2008.06499, toappear} {and the parent effective field theory that describes the phase transition and neighborhood phases}, 
such as those explored in 4d \cite{Wan2018djlW2.1812.11955, Anber2018iof1805.12290, Cordova2018acb1806.09592DumitrescuClay, BiSenthil1808.07465, WangYouZheng1910.14664}. {The underlying mathematical structure suggests a 4d version of particle-vortex duality, S-duality, T-duality, or mirror symmetry.}
See \Fig{fig:Fig6UU}, \Fig{fig:Fig7}, and  \Fig{fig:Fig8} for illustrations. 

\subsection{Summary of Ultra Unification and Quantum Matter in Drawings}
\label{label:Summary}

Let us summarize what we have done in this work in drawings. 

\noindent
$\bullet$ \Fig{fig:Fig2SM}:
We have started from the Nature given Standard Model (SM) quarks and leptons, and their quantum numbers (see Tables in \cite{JW2006.16996}), in three generations. 

\noindent
$\bullet$ \Fig{fig:Fig3GUT}:  
We have included gauge forces and various Higgs for SM and Grand Unification (GUT).

\noindent
$\bullet$ \Fig{fig:Fig4extra}:
{After the essential check of the {anomaly and cobordism constraints}, the detection of the $\Z_{16}$ global anomaly for 15n Weyl fermion SM and GUT implies 
that we can choose (as one of many options)
to realize our 4d world living on an extra-dimensional 5d invertible TQFT (a 5d topological superconductor, mathematically a 5d cobordism invariant).
Here \Fig{fig:Fig4extra} may be understood as a one-brane 4d world with an extra-large fifth dimension.}

\noindent
$\bullet$ \Fig{fig:Fig5UU}: {{\bf Ultra Unification} (UU) incorporates the {SM, GUT, and Topological forces into the same theory}
(that this theory can also be coupled to curved spacetime geometry and gravity in a background non-dynamical way).
(1) The upper left has the 4d SM and GUT. (2) The upper right has the 4d non-invertible TQFT. 
(3) The bottom has the 5d invertible TQFT (alternatively  5d non-invertible TQFT if we dynamically gauge the full $\Z_{4,X}$).
The three sectors can communicate with each other mediated via the dynamical $\Z_{4,X}$ gauge forces.
Here \Fig{fig:Fig5UU} may be understood as a multi-branes or two-brane 4d world with an extra fifth dimension.}
The issues of mirror fermion doubling \cite{NN8119} on the mirror world, depending on the precise anomaly index on the mirror sector,
may be fully trivially gapped (if anomaly-free), {may contain a mirror chiral gauge theory or unparticle conformal field theory, or may be topological order gapped with a low energy TQFT}.
The issues of gapping mirror fermions 
are tackled in the past starting from Eichten-Preskill \cite{Eichten1985ftPreskill1986} and by many recent works 
\cite{Wen2013oza1303.1803, Wen2013ppa1305.1045, Wang2013ytaJW1307.7480, 
You2014oaaYouBenTovXu1402.4151,
YX14124784, BenTov2015graZee1505.04312, Kikukawa2017gvk1710.11101, Kikukawa2017ngf1710.11618, Wang2018ugfJW1807.05998, WangWen2018cai1809.11171, RazamatTong2009.05037, Catterall2020fep}.

\noindent
$\bullet$ \Fig{fig:Fig6UU}:  In Quantum Matter terminology, we show that SM and GUT belong to a framework of 
a continuous gauge field theory, Anderson-Higgs (global or gauge) symmetry-breaking mass, and Ginzburg-Landau paradigm. In contrast, 
the new sectors that we introduce are beyond Ginzburg-Landau paradigm.
The new sectors include a fermionic discrete gauge theory,
symmetry-extension topological mass, and modern issues on symmetry, topology, nonperturbative interactions, and short/long-range entanglements. 

\noindent
$\bullet$ \Fig{fig:Fig7} and \Fig{fig:Fig8}: In general, our SM vacuum (with possible BSM corrections that we denote SM$^*$)
may live in a landscape of quantum vacua (e.g., see recent works \cite{WangYou2106.16248, WangYou2111.10369GEQC}). 
Possible quantum vacua tuning parameters may be the GUT-Higgs potential 
or other QFT parameters that can induce quantum phase transitions from SM to neighbor GUT vacua 
in an SM deformation class \cite{NSeiberg-Strings-2019-talk, WangWanYou2112.14765}.    
UU may not only be just a higher-energy effective field theory of SM$^*$,
but also provide a parent effective field theory to go between SM, SM$^*$, or different GUT vacua,
from the left-hand sided 15n-Weyl-fermion model plus a 4d TQFT, a 4d CFT, or a 5d iTQFT,
to the right-hand sided 16n-Weyl-fermion model in 4d, via a topological quantum phase transition
through an energy-gap-closing gapless quantum critical region.

\noindent
$\bullet$ \Fig{fig:Fig9}:
{UU provides alternative new ways to explain the mass of the neutrinos. 
The traditional seesaw mechanism argues that the right-handed sterile neutrinos have a \emph{Majorana mass} $m_{\rm M}$,
while the left-handed paired with right-handed neutrinos get a \emph{Dirac mass} $m_{\rm D}$; 
the mass eigenstate allows a small mass $\frac{m_{\rm D}^2}{m_{\rm M}} \ll {m_{\rm D}}$; these quadratic masses break the $\Z_{4,X}$ symmetry.\\
In contrast, UU replaces some of sterile neutrinos with a TQFT sector within the same anomaly index. 
TQFT energy gap $\Delta_{\rm TQFT}$ is the \emph{topological mass} gap.
The $\Delta_{\rm TQFT}$ may sit at any energy scale below GUT or Planck scale $M_{\rm GUT,Pl}$.
The symmetric TQFT can preserve $\Z_{4,X}$ but still give a topological mass; 
the TQFT may also have  
$\Z_{4,X}$-topological defects which trap the zero modes.
(Proliferating the topological defects restores the $\Z_{4,X}$-symmetry driving to the symmetric TQFT phase.)
The nearly gapless \emph{left-handed neutrinos} ($\nu_{e,L},\nu_{\mu,L},\nu_{\tau,L}$) travel in waves and interact with 
TQFT defect's \emph{zero modes}. 
Quantum interference between {left-handed neutrinos} and zero modes
 possibly causes \emph{neutrino oscillations}.
The energy spectrum near the defect (e.g., vortex points/strings) has some energy subgap 
$\Delta_{\rm sub} \lesssim \frac{\Delta_{\rm TQFT}^2}{M_{\rm GUT,Pl}}$ above the zero energy modes.\footnote{This 
dimensional analysis
$\Delta_{\rm sub} \lesssim \frac{\Delta_{\rm TQFT}^2}{M_{\rm GUT,Pl}}$
is in analogy with the vortex energy subgap $\Delta_{\rm sub} \simeq \frac{\Delta_{\rm SC}^2}{E_F}$ 
of superconductor gap $\Delta_{\rm SC}$ and Fermi energy $E_F$
in the finite-density fermionic superconductor system. This idea is reported by the author 
in the Harvard String Lunch seminar (December 4, 2020).} 
This interacting mechanism $\Delta_{\rm sub} \lesssim \frac{\Delta_{\rm TQFT}^2}{M_{\rm GUT,Pl}} \ll {m_{\rm D}}$
(different from the traditional seesaw mechanism $\frac{m_{\rm D}^2}{m_{\rm M}} \ll {m_{\rm D}}$)
may also give the left-handed neutrinos small masses.}

\noindent
Of course, there could also be mixed scenarios (enumerated in \Sec{sec:Consequences})
such that we have some right-handed neutrino sector and some TQFT sector,
so both sectors can contribute the small $\frac{m_{\rm D}^2}{m_{\rm M}}$
and small $\Delta_{\rm sub} \lesssim \frac{\Delta_{\rm TQFT}^2}{M_{\rm GUT,Pl}}$ respectively 
to the mass origin of the left-handed neutrinos.

\newpage
       
\begin{figure}[h!] 
  \centering
\hspace{-1.cm}
  \;\includegraphics[width=6.6in]{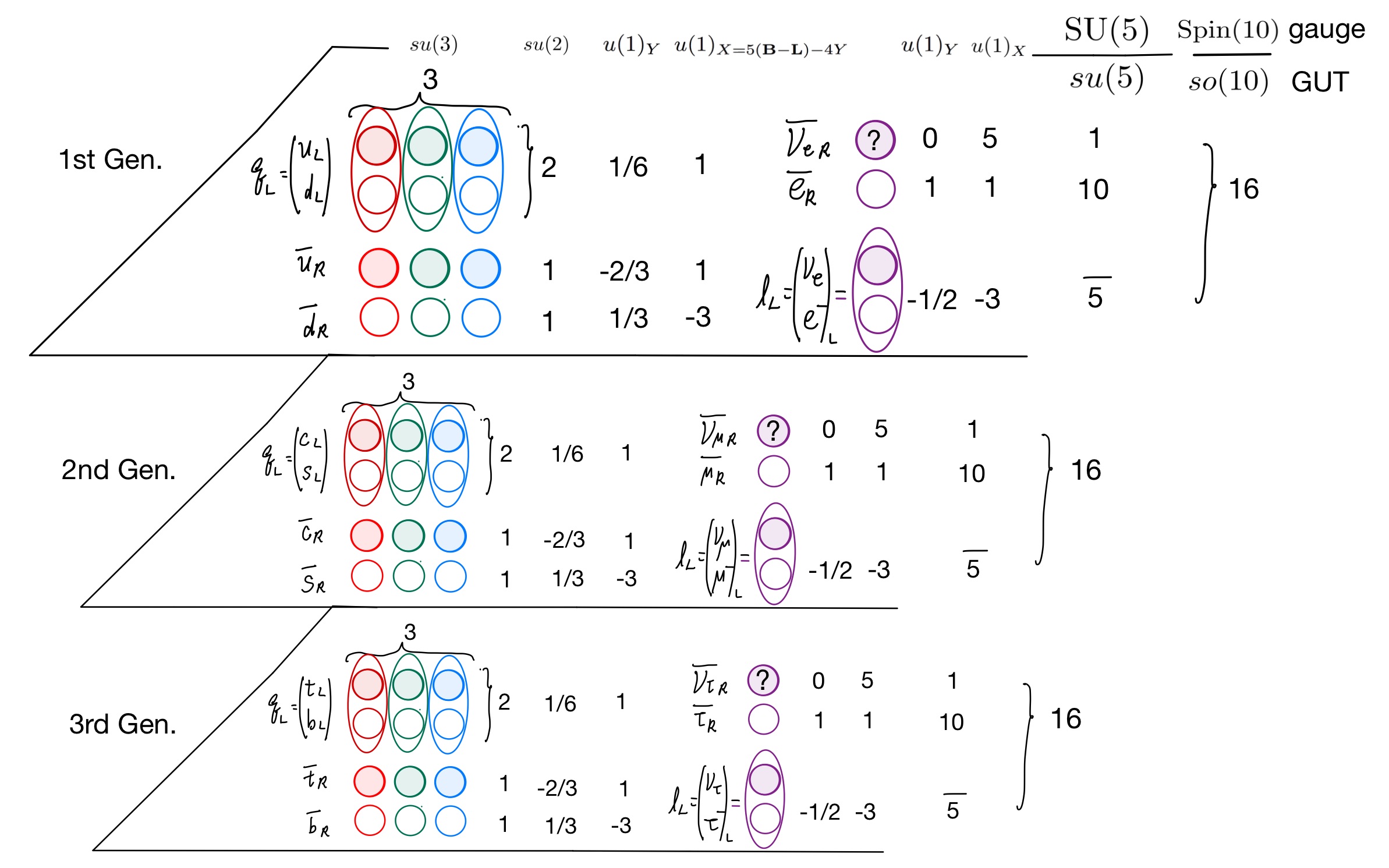}
  \caption{Standard Model, the $su(5)$ GUT and the $so(10)$ GUT quarks and leptons. {Each Weyl fermion, represented by a color circle, 
  is a Lorentz spinor ${\bf 2}_L$ under the
  spacetime symmetry group ${{\Spin(1,3)}}$.}}
  \label{fig:Fig2SM}
\end{figure}


\begin{figure}[h!] 
  \centering
  \;\includegraphics[width=6.5in]{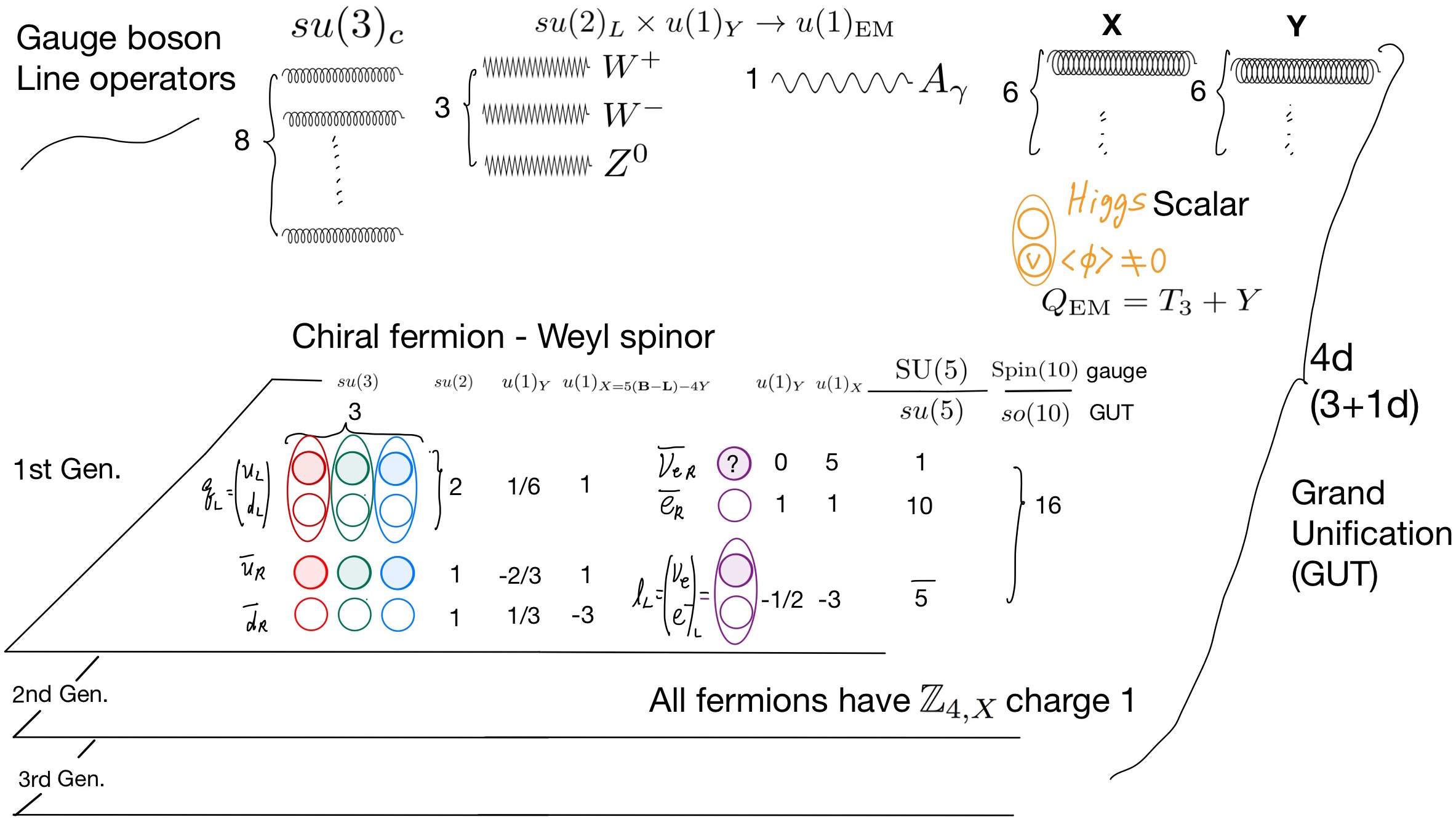}
  \caption{Include gauge forces and (various) Higgs to the
  Standard Model and GUT. Other heavier excitations 
  such as a magnetic monopole would occur as the open end of 't Hooft (world-)line operator.}
  \label{fig:Fig3GUT}
\end{figure}

\newpage

\begin{figure}[h!] 
  \centering
  \;\includegraphics[width=5.4in]{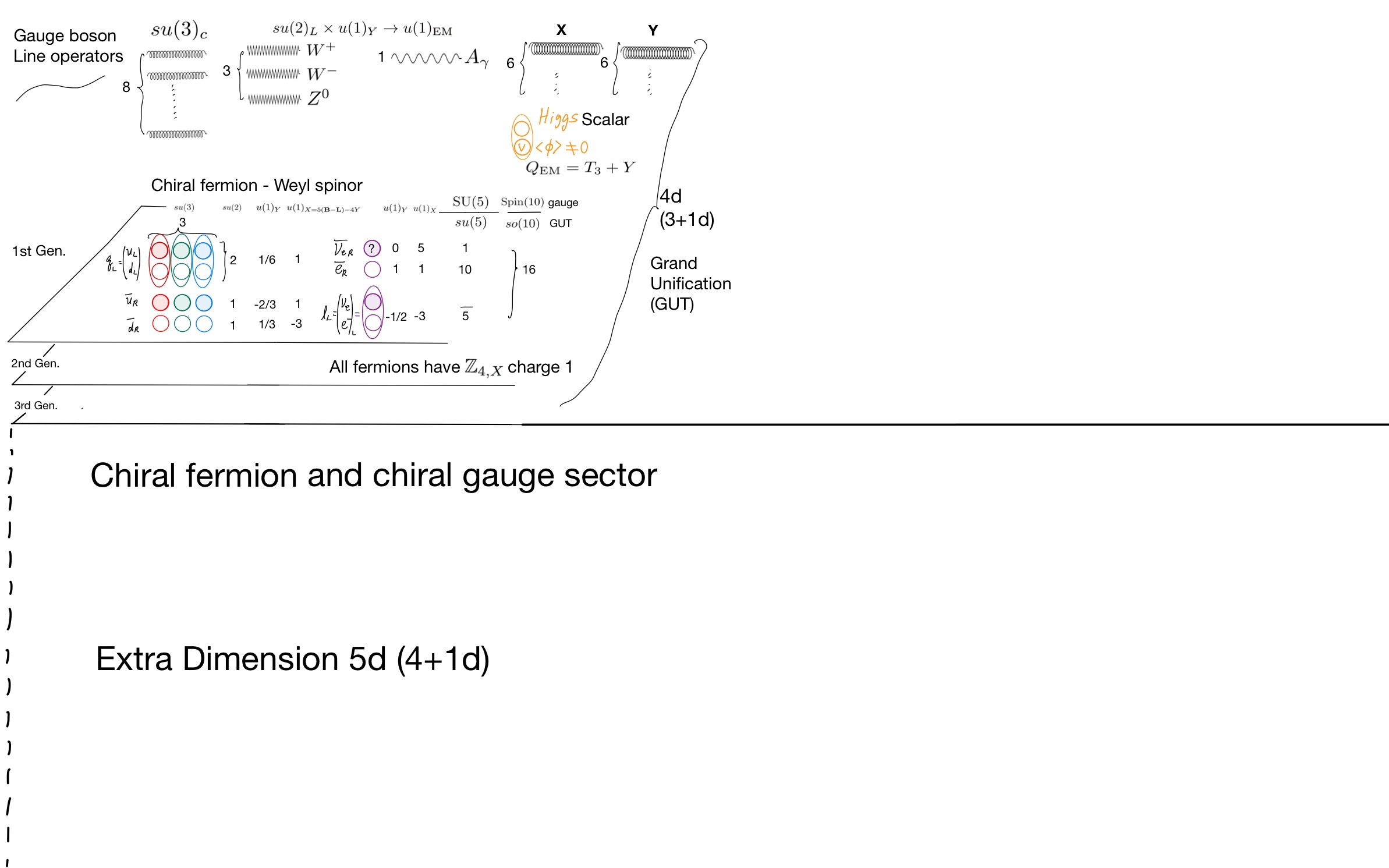}
  \caption{Detection of the $\mathbb{Z}_{16}$ global anomaly for 15n Weyl fermion SM and GUT.
  It can be realized as adding an extra-dimensional 5d invertible TQFT (i.e., a 5d cobordism invariant), where the
  4d world lives on its boundary.
    }
  \label{fig:Fig4extra}
\end{figure}


\begin{figure}[h!] 
  \centering
  \;\includegraphics[width=6.26in]{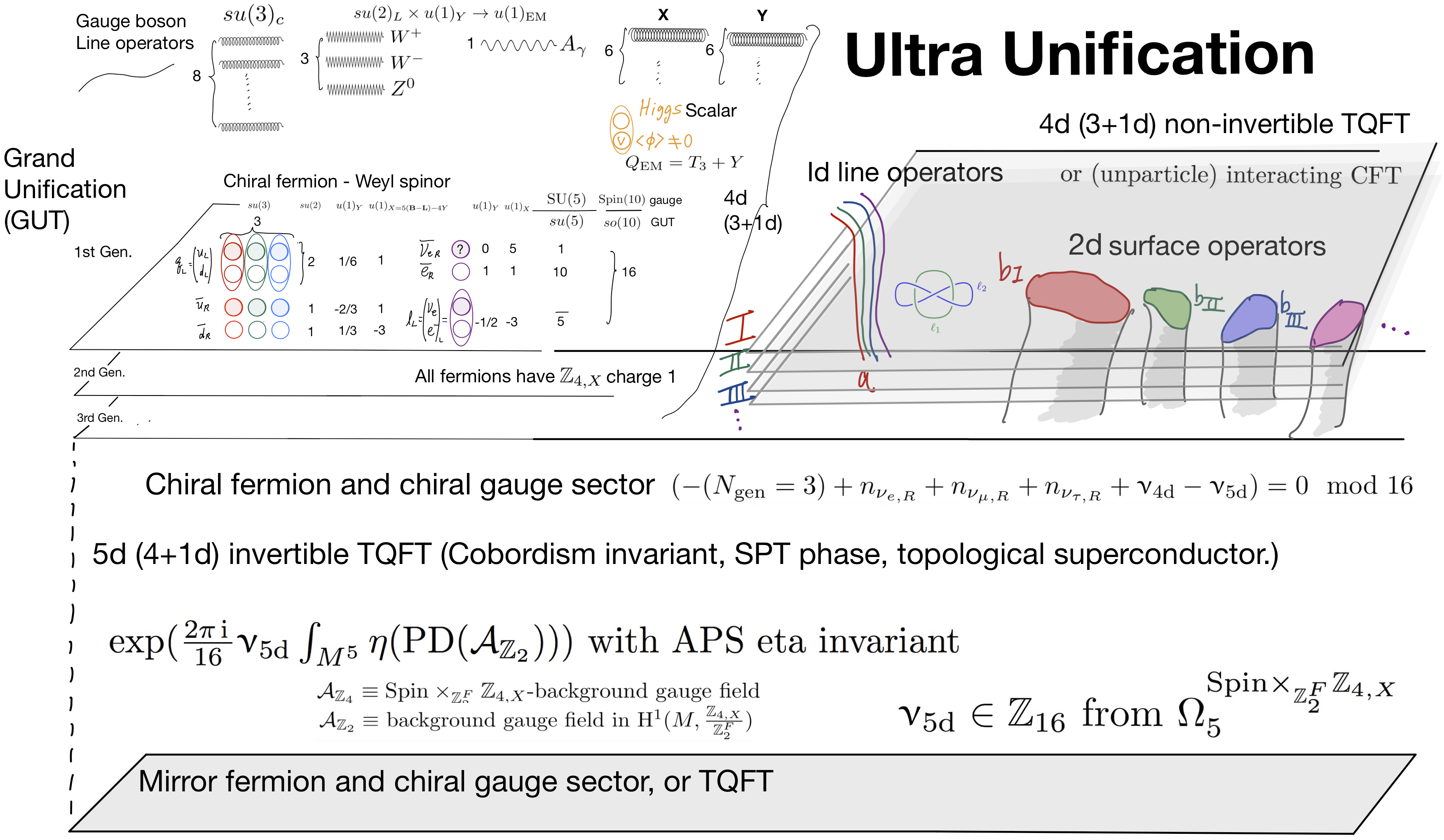}
  \caption{{\bf Ultra Unification} incorporates the {\bf Strong, Electromagnetic, Weak, GUT, and Topological Forces into the same theory}
in the QFT/TQFT path integral 
(that this theory can also be coupled to curved spacetime geometry and gravity in a background non-dynamical way).
(1) The upper left has the 4d SM and GUT. (2) The upper right has the 4d non-invertible TQFT. 
(3) The bottom has the 5d invertible TQFT (alternatively  5d non-invertible TQFT if we dynamically gauge the $\mathbb{Z}_{4,X}$ or ${\mathbb{Z}_{4,X}}/{\Z_2^F}$). 
The three sectors can communicate with each other mediated via the {dynamical 
discrete gauge forces}.
    }
  \label{fig:Fig5UU}
\end{figure}

\newpage

\begin{figure}[h!] 
  \centering
\hspace{-1.15cm}
  \includegraphics[width=6.in]{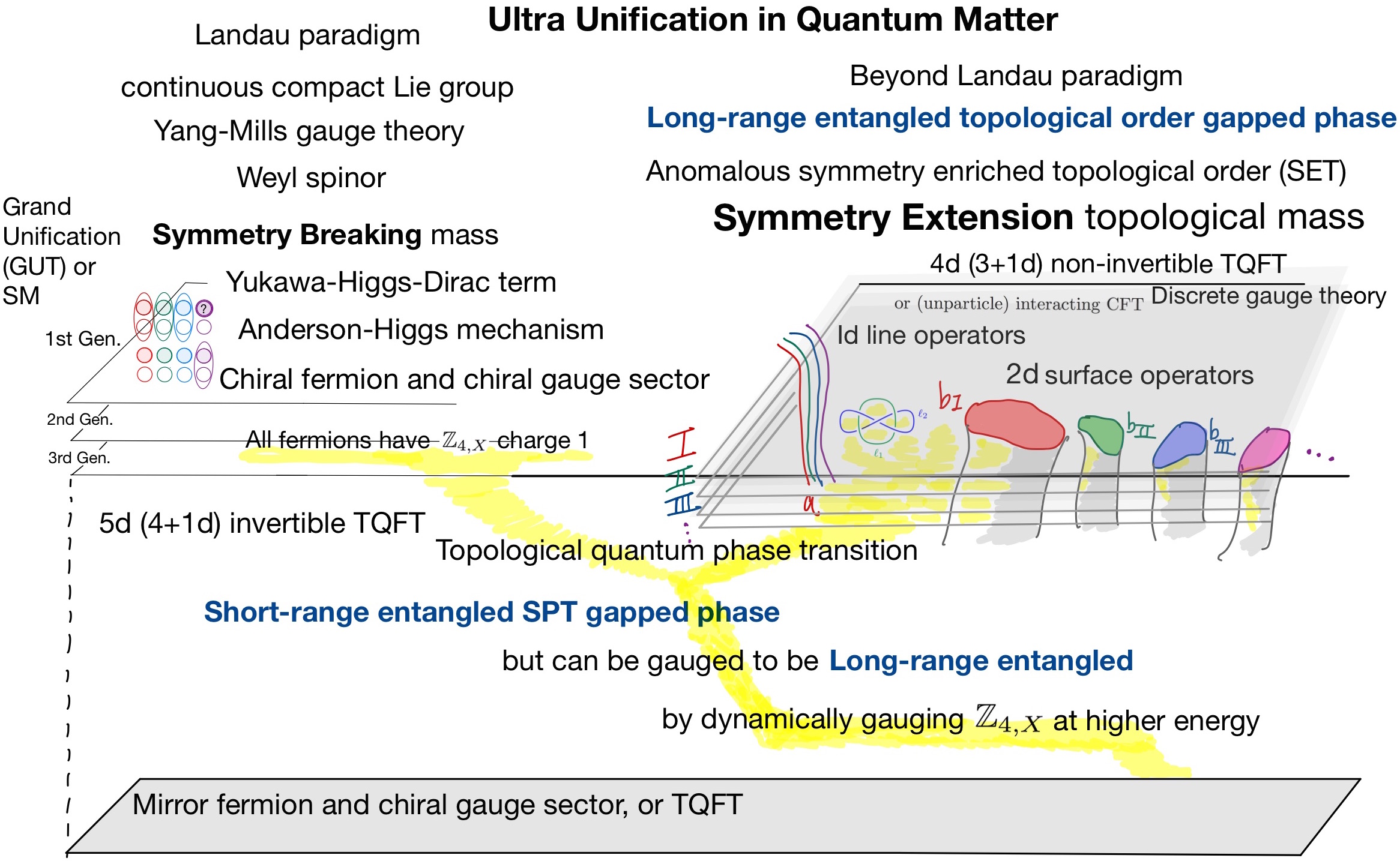}\;
  \caption{{\bf Ultra Unification} summarized in terms of Quantum Matter terminology.
  We can trade some even number of right-handed neutrinos (the 16th Weyl fermions, 
  possibly with \emph{symmetry-breaking} masses)
  for the \emph{symmetry-extension} topological order sector in 3+1d, via a topological quantum phase transition.
  {Once the 
  $\Z_{4,X}$ is dynamically gauged (e.g., at higher energy), the 3+1d theory and 4+1d bulk all are coupled and correlated with each other via the
 topological $\Z_{4,X}$-gauge force. In a colloquial sense, our Standard Model world may live with the neighbors of 3+1d \emph{intrinsic topological order} or 3+1d \emph{unparticle CFTs}, and also live on the boundary of some medium of 4+1d \emph{topological quantum computer}.}
    }
  \label{fig:Fig6UU}
\end{figure}

\begin{figure}[h!] 
  \centering
\hspace{-1.35cm}
\includegraphics[width=6.in]{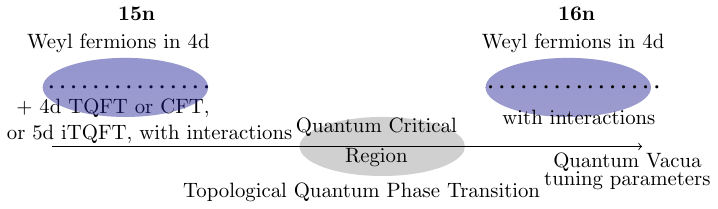} 
\caption{%
Ultra Unification contains all pertinent low-energy scenarios (enumerated in \Sec{sec:Consequences}), 
such that we can have a topological quantum phase transition
from the left-hand sided 15n-Weyl-fermion model plus 4d TQFT/CFT or 5d iTQFT,
to the right-hand sided 16n-Weyl-fermion model in 4d (shown in two shaded blue regions as two different phases of quantum vacua 
with different low-energy and different ground state sectors), 
tuning through a \emph{gapless quantum critical region} (shown as the schematic gray region).
The 15 dots and 16 dots on the left and right regions represent the number of Weyl fermions. 
The horizontal axis indicates the possible quantum vacua tuning parameter(s).
Quantum phase transition and quantum criticality may also exhibit supersymmetry phenomena.
}
 \label{fig:Fig7}
\end{figure}

\begin{figure}[h!] 
  \centering
\hspace{-1.35cm}
\includegraphics[width=4.5in]{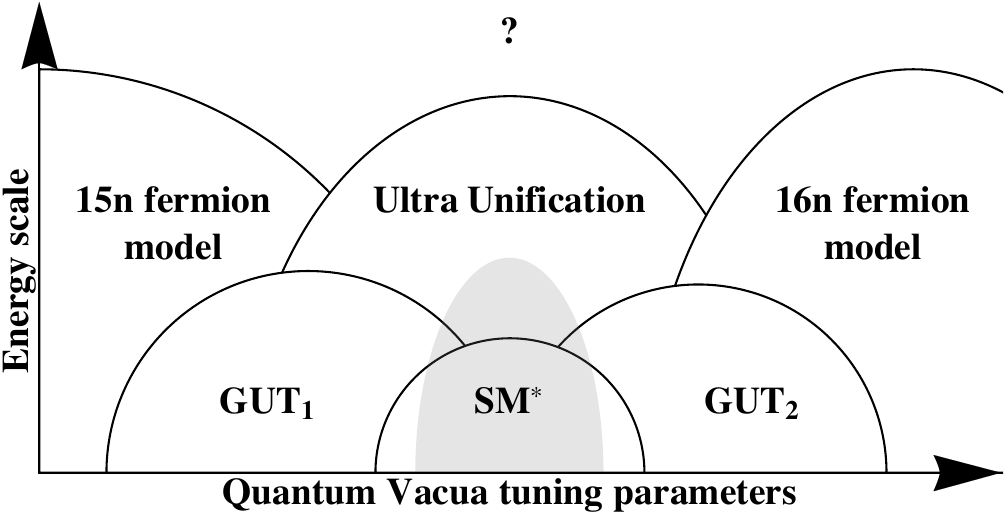}\;\quad 
\caption{SM vacuum (with possible BSM corrections denote as SM$^*$)
may live in a landscape of quantum vacua in a quantum phase diagram (e.g., see recent works \cite{WangYou2106.16248, WangYou2111.10369GEQC}). 
Possible quantum vacua tuning parameters may be the GUT-Higgs potential 
or other QFT deformation parameters that can induce quantum phase transition from SM to neighbor GUT vacua 
in an SM deformation class \cite{NSeiberg-Strings-2019-talk, WangWanYou2112.14765}.    
In \Refe{WangYou2106.16248, WangYou2111.10369GEQC}'s viewpoint,
Ultra Unification may not only be just a higher-energy effective field theory of SM$^*$,
but also provide a parent effective field theory to go between SM, SM$^*$, or different GUT vacua, 
from the left-hand sided 15n-Weyl-fermion model plus 4d TQFT/CFT or 5d iTQFT,
to the right-hand sided 16n-Weyl-fermion model in 4d, 
tuning through a \emph{gapless topological quantum critical region} (the schematic gray region).
}
 \label{fig:Fig8}
\end{figure}

\begin{figure}[h!] 
  \centering
\includegraphics[width=6.55in]{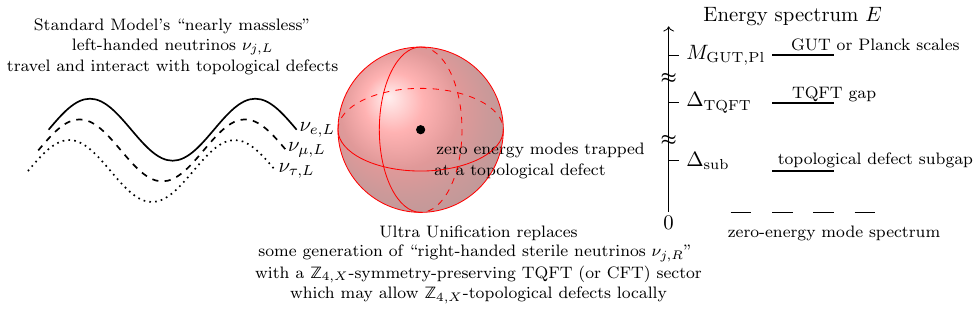}
\caption{Ultra Unification (UU) also provides alternative new possible ways to explain the mass of the neutrinos. 
The traditional seesaw mechanism argues that the right-handed sterile neutrinos have a \emph{Majorana mass} $m_{\rm M}$,
while the left-handed paired with right-handed neutrinos get a \emph{Dirac mass} $m_{\rm D}$; 
the mass eigenstate allows a small mass $\frac{m_{\rm D}^2}{m_{\rm M}} \ll {m_{\rm D}}$; these quadratic masses break the $\Z_{4,X}$ symmetry.
In contrast, UU replaces some of sterile neutrinos to a TQFT sector with the same anomaly index, which can have a \emph{topological mass} of TQFT energy gap $\Delta_{\rm TQFT}$.
The $\Delta_{\rm TQFT}$ may sit at any energy scale below GUT or Planck scale $M_{\rm GUT,Pl}$.
The symmetric TQFT can preserve $\Z_{4,X}$ but still give a topological mass; 
the TQFT may also have  
$\Z_{4,X}$-topological defects which trap the zero modes.
(Proliferating the topological defects restores the $\Z_{4,X}$-symmetry driving to the symmetric TQFT phase.)
The nearly gapless \emph{left-handed neutrinos} ($\nu_{e,L},\nu_{\mu,L},\nu_{\tau,L}$) travel in waves and interact with 
TQFT defect's \emph{zero modes}, which quantum interference possibly causes \emph{neutrino oscillations}.
The energy spectrum near the defect (such as vortex points/strings) has energy subgap 
$\Delta_{\rm sub} \lesssim \frac{\Delta_{\rm TQFT}^2}{M_{\rm GUT,Pl}}$, 
in analogy with the vortex subgap $\Delta_{\rm sub} \simeq \frac{\Delta_{\rm SC}^2}{E_F}$ of superconductor gap $\Delta_{\rm SC}$ and Fermi energy $E_F$.
In contrast to the traditional seesaw mechanism $\frac{m_{\rm D}^2}{m_{\rm M}} \ll {m_{\rm D}}$,
this interacting mechanism $\Delta_{\rm sub} \lesssim \frac{\Delta_{\rm TQFT}^2}{M_{\rm GUT,Pl}} \ll {m_{\rm D}}$
may also give the left-handed neutrinos small masses.}
 \label{fig:Fig9}
\end{figure}

\newpage
\section{Acknowledgements and Bibliography} 
JW thanks the participants of Quantum Matter in Mathematics and Physics program at
Harvard University CMSA for the enlightening atmosphere.
Part of this work had been presented at Higher Structures and Field Theory at Erwin Schr\"odinger Institute in Wien (August 4, 2020) \cite{ESIJWUltraUnification},
and at Harvard University Particle Physics Lunch seminar (November 30, 2020) and String Lunch seminar (December 4, 2020). 
JW thanks the valuable feedbacks from the seminar participants.\footnote{Instead of writing or drawing an image of the author's mental feelings, a piece of  
Johann Sebastian Bach's  music ``The Goldberg Variations, BWV 988 (1861)'' (e.g., Variatio 25. a 2 Clav.: Adagio) may illuminate this well.} 
%
This work is supported by 
NSF Grant DMS-1607871 ``Analysis, Geometry and Mathematical Physics'' 
and Center for Mathematical Sciences and Applications at Harvard University.


\bibliographystyle{Yang-Mills}
\bibliography{BSM-SU3SU2U1-cobordism-UU.bib}

 \end{document}